# Chemistry by Design (CbyD): A Machine-Readable Dataset of Expert-Curated Total Synthesis Routes

*Ryan E. Williams[1,†], Suong B. A. Tran[2,†], Hayden M. Leatherwood[1], Jihye Roh[3], Aquila Wolff[3], John G. Federice[1], Nicholas R. Lauta[1], Kevin Yu[4], Steven J. Nick[1], Zlatko Jončev[5], Philippe Schwaller[5], Connor W. Coley[3,4,*], Jon T. Njardarson[1,*]*

[1]Department of Chemistry and Biochemistry, University of Arizona, Tucson, AZ, USA

[2]Department of Chemistry, Massachusetts Institute of Technology, Cambridge, MA, USA

[3]Department of Chemical Engineering, Massachusetts Institute of Technology, Cambridge, MA, USA

[4]Department of Electrical Engineering and Computer Science, Massachusetts Institute of Technology, Cambridge, MA, USA

[5]Institute of Chemical Sciences and Engineering, École Polytechnique Fédérale de Lausanne, Lausanne, Switzerland

[†]These authors contributed equally.

*Email: ccoley@mit.edu, njardars@arizona.edu

**ABSTRACT:** Chemistry by Design (CbyD) is an open-access educational database of published total syntheses launched in 2011. It was designed to help students and practitioners learn organic reactions and synthetic strategies by depicting every step of published natural product and drug syntheses. Motivated by advances in machine learning for organic synthesis and by requests from academic and industrial researchers, we present a machine-readable version of CbyD for use in cheminformatics and computer-aided synthesis planning (CASP). The released dataset (v2026.11) contains 3,502 routes comprising 56,640 reactions toward 2,902 unique targets. Compared with patent-derived datasets, CbyD reactions are more complex, more frequently change stereochemistry, and are less likely to be covered by patent-derived reaction templates, and CbyD routes are longer and more convergent than those in commonly used CASP benchmarks. We also analyze how molecular complexity changes along these routes and identify transformations associated with large complexity increases. The CbyD dataset will be expanded through versioned public releases that incorporate new routes and community-contributed corrections.

## INTRODUCTION

In early 2010, the Njardarson group began developing Chemistry by Design (CbyD), an educational resource for learning organic reactions and synthetic strategies through published syntheses of natural products and drugs.[1] CbyD presents synthetic routes primarily through chemical structure drawings, making the content largely independent of language and accessible to a broad audience. An interactive flash-card mode allows users to test their knowledge by predicting reagents and products in the forward direction, or reagents and starting materials when navigating a route in reverse. Typical synthesis publications condense multistep sequences and omit the preparation of starting materials, leaving readers to infer what is not shown. In contrast, CbyD shows every starting material, reagent, and product for every step, so that users can follow each design and reaction decision made in the original synthesis.

The CbyD website (chemistrybydesign.oia.arizona.edu) was launched in June 2011, followed a few months later by a free mobile application for iOS and Android devices. Expert-curated content has been added continuously since then, making CbyD the largest freely accessible collections of natural product total syntheses, along with a substantial collection of drug syntheses.

Beyond its educational use, this large CbyD collection of drug and total syntheses is potentially valuable for computer-aided synthesis planning (CASP), particularly for multistep retrosynthetic planning. The large datasets currently used in CASP consist predominantly of one-step reactions, such as USPTO,[2] Pistachio,[3] Reaxys,[4] and SciFinder[5].[6,7] Existing route-level datasets are instead reconstructed almost exclusively from patent reaction records by linking the product of one reaction to a reactant of another.[8,9,10] These reconstructed routes can contain sequences that would not be executed in practice, such as an esterification and hydrolysis that return to the same intermediate, or consecutive protection and deprotection steps, as shown with examples in **Section S4.1**. In addition, patent reactions are generally not subject to scientific peer review, and the reported conditions may not represent optimized procedures. While the SciFinder and Reaxys datasets have a more representative mixture of patent and literature-published reactions, there are significant restrictions on their access and use. The USPTO-derived and Pistachio datasets are much more frequently adopted for cheminformatics and CASP, despite being narrower in information sources.

Open access, machine-readable datasets of complete, experimentally executed synthesis routes remain limited.[11] However, CbyD (4356 syntheses as of November 2026 and growing weekly),[1] Reich's collection (669 syntheses),[12] and SynArchive (323 syntheses)[13] contain experimentally reported synthesis routes, yet only present reactions and conditions as images. CbyD is not only the largest of these collections but is also the only one that also records all reaction steps used to assemble a target structure and not only the longest linear sequence.[14]

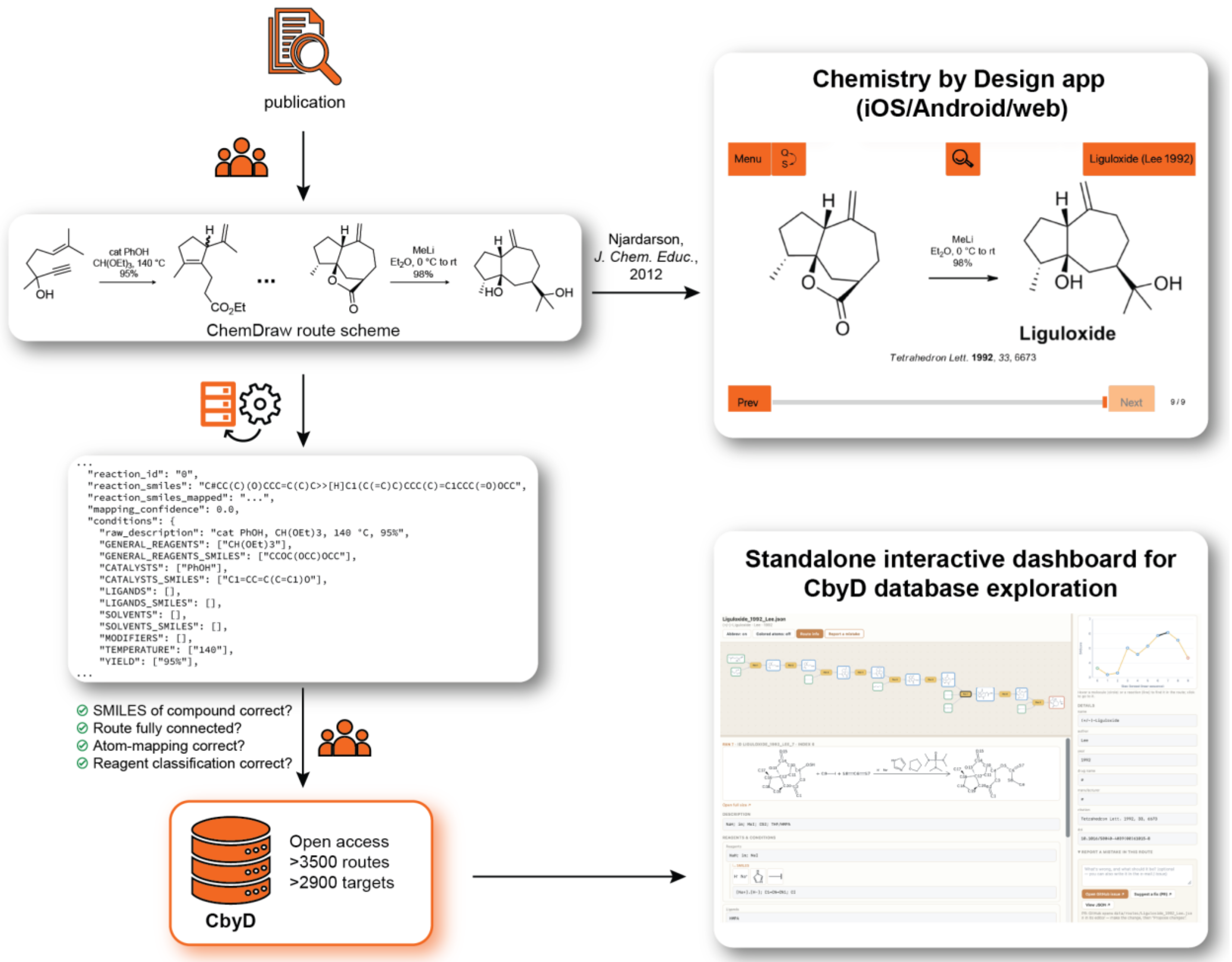

**Figure 1.** Overview of the CbyD curation process. Published syntheses are redrawn as ChemDraw route schemes showing every step. These schemes are converted into structured records with reaction SMILES, parsed conditions, and atom mappings. The records are then manually checked before being added to the open-access CbyD dataset. A route visualization dashboard is available for browsing the structured data, analogous to the official CbyD website.

To address this gap, we digitized the CbyD collection, building on the expertise of its original curators and contributors. We converted the original ChemDraw files into structured records and curated them to preserve and standardize the source information (**Fig. 1**). This conversion proved considerably more difficult than anticipated, largely because chemical information is drawn inconsistently across the literature. For example, abbreviations for reagents, solvents, and other species vary widely across the original CbyD ChemDraw files, with the same chemical species often represented by different names or abbreviations. To standardize these entries, we developed and manually curated abbreviation dictionaries comprising over 9,000 SMILES definitions. Structures drawn in perspective also frequently caused errors in structure extraction and stereocenter assignment resulting in redrawing of multiple structures and even large sections of sequences as the only solution to realize a successful extraction.

In this work, we describe the conversion and curation of CbyD into a machine-readable dataset comprising total syntheses of natural products and approved drugs. The result of this multi-year effort is an open-access CbyD dataset in which each reaction is annotated with SMILES, parsed reaction conditions, and atom mappings. During the final stages of this work, AbSynth was released as a structured synthetic route dataset.[15] We observed substantial overlap between its content and the routes available in the curated CbyD educational website (**Section S4.2**). CbyD differs in providing atom-mapped reactions with classified reagents. The dataset is openly available without registration and is accompanied by a standalone, user-friendly dashboard that allows users to browse and visualize complete synthesis routes and their associated reaction information. The 2026.11 version of our dataset contains 3,502 multistep routes comprising 56,640 one-step reactions. We characterize the dataset at the reaction and route levels and show that, compared with patent-derived datasets, CbyD reactions more frequently change stereochemistry and fall outside the coverage of patent-derived templates, and CbyD routes are longer and more convergent. We anticipate that these machine-readable representations of experimentally executed syntheses will be useful for research in cheminformatics and CASP.

## DATASET CURATION

CbyD was originally developed as an educational resource to present complete synthetic routes, with every reaction step shown explicitly. To achieve this, each synthesis was manually recreated in ChemDraw,[16] a widely used chemical drawing program among synthetic organic chemists. These files were then rendered as individual images of each starting material, reaction condition (including reagents, solvents, and temperature, where reported), and product for the CbyD website and application. As the original format was developed in 2011 without consideration for machine readability, the ChemDraw files served as the sole source for conversion into a machine-readable dataset.

### Conversion of the original ChemDraw files into machine-readable format.

Each ChemDraw scheme is converted into a structured JSON file by parsing the CDXML or CDX file and extracting molecular structures and reaction condition text. Each reaction arrow defines one reaction step, and structures are assigned to the reactant, product, or agent fields based on their position relative to the arrow. Molecular structures are converted to RDKit[17] molecules directly from their CDXML representation. Condition text above the reaction arrow is parsed and classified into reagents, catalysts, ligands, solvents, modifiers, temperatures, and yields. The complete reaction is then represented as SMILES, with corresponding atoms in the reactants and products (i.e., atom-mapping) identified using RXNMapper.[18] Because the original drawings occasionally contain ambiguous or unsupported ChemDraw objects, the conversion process includes automated checks for unresolved structures, missing reactants or products, and other extraction failures (**Section S1**).

Extracted records were also manually inspected and corrected, because many errors arise from how a synthetic scheme was originally drawn rather than from the extractor itself. For example, bridgehead stereocenters drawn in perspective or chiral centers without directional wedges or hashes carry no stereochemical information, and older versions of ChemDraw and RDKit cannot perceive chirality from ring geometry alone. Even when directional wedges or hashes are present, the assigned configuration can change with the drawn perspective of the ring system (**Section S1.3.2**). Abbreviated structures and legacy text encodings caused similar failures. These cases were identified, re-extracted, and verified manually. Details of the extraction process are described in **Section S1.1**.

**Data structure.**

The data format is described in **Fig. 2**; each synthesis route is stored as a single JSON file. We also provide a tree-structured representation of each route, in which the target is the root node and molecule and reaction nodes are linked by parent–child relationships. These trees are constructed automatically by connecting each reactant to the reaction that produces it, with additional considerations specific to CbyD (**Section S1.7**).

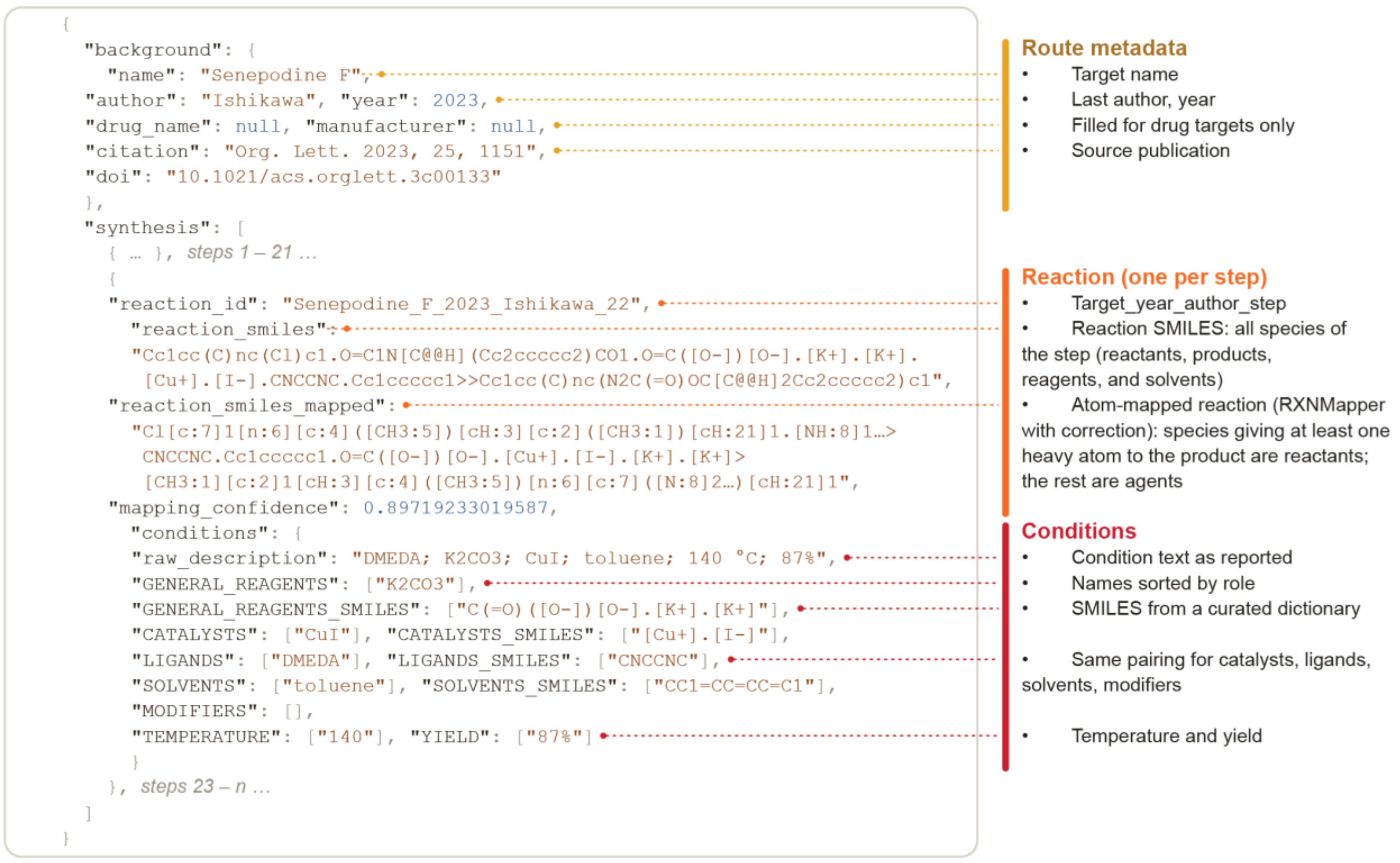

**Figure 2.** Hierarchical data structure of a CbyD route record.

### Dataset availability and versioning.

The CbyD dataset is released as an open-access resource at [https://github.com/chemistry-by-design/cbyd](https://github.com/chemistry-by-design/cbyd) under the CC BY 4.0 license. Each public release will be a tagged snapshot of the dataset at a fixed point in time, so that any released version can be cited and reproduced exactly. Because the collection will keep growing and may require corrections to existing records, releases follow calendar versioning of the form CbyD vYYYY.MM. This first release, CbyD v2026.11, contains 3,502 multistep routes comprising 56,640 individual one-step reactions. These routes cover 2,902 unique target molecules, including 2,437 natural products and 465 pharmaceuticals.

We welcome contributions from the community. The dataset is distributed with a standalone HTML dashboard for route visualization. From any route, users can report an error by email or open a pull request against the corresponding record. Based on our experience digitizing CbyD, we also recommend that journals publishing total syntheses encourage authors to provide ChemDraw files or structured data of their synthesis routes, ideally following the submission guidelines available in our GitHub repository. All proposed additions and corrections are manually reviewed before being incorporated into the curated dataset.

## RESULTS AND DISCUSSION

### CbyD contains a diverse collection of synthetic routes and target molecules.

CbyD is curated primarily from leading synthetic organic chemistry journals and therefore emphasizes total syntheses of complex natural products and pharmaceuticals. The number of reported routes increases substantially after the year 2000 and continues to rise through the most recent publications (**Fig. 3A and Table S1**). Most targets in CbyD have only one reported route, although multiple routes are available for a substantial number of targets. Some frequently synthesized natural products, including englerin A, strychnine, morphine, aspidospermidine, and hirsutene, have more than fifteen reported routes each (**Fig. 3B and Table S2**). Multiple routes to the same target allow direct comparison of how disconnection choice, reaction selection, and step order shape the overall route. They also serve as a unique set of literature-validated answers that better reflects the one-to-many nature of retrosynthesis that tends to be overlooked in CASP evaluation.

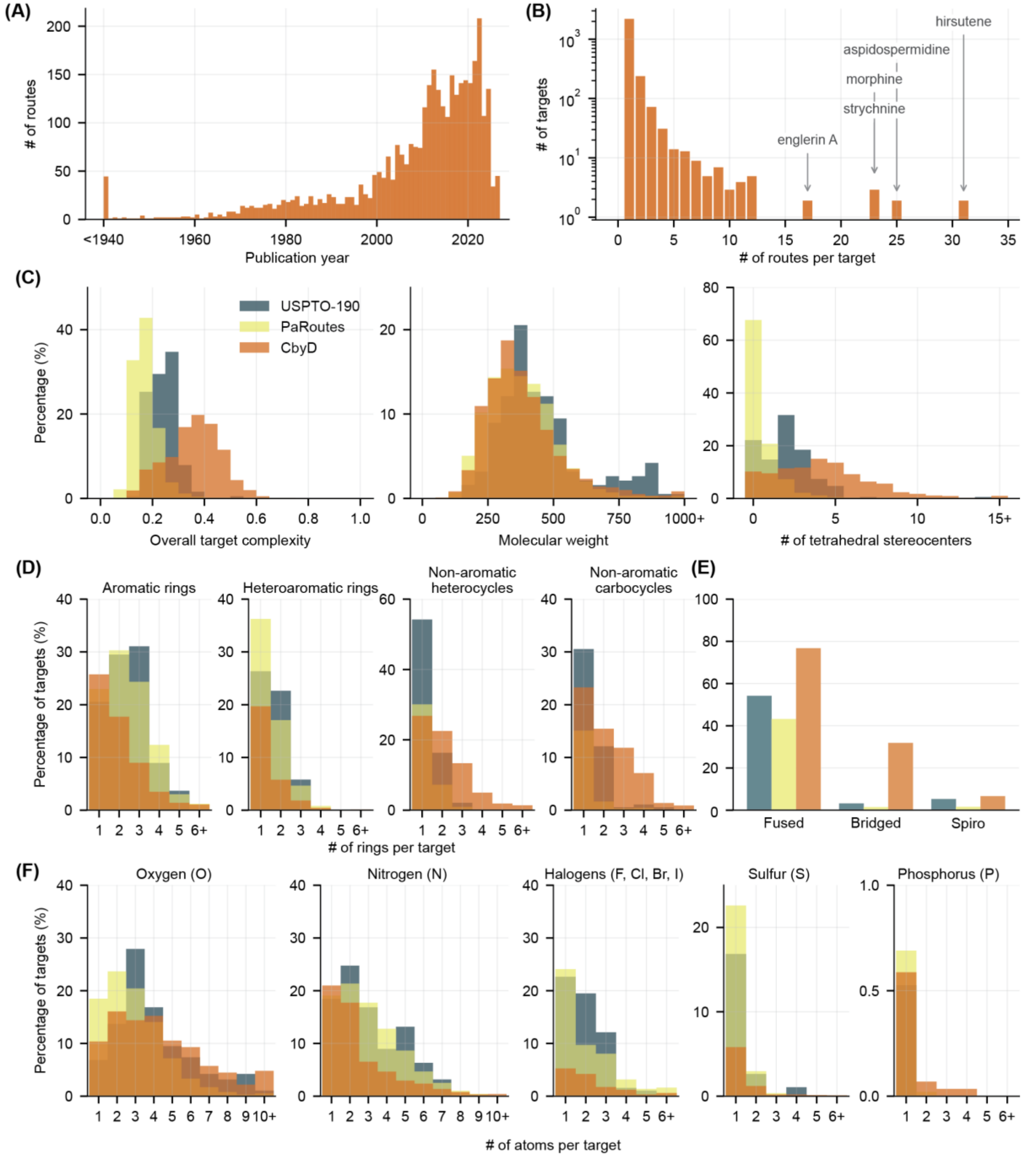

**Figure 3. (A)** Distribution of publication years for the targets. **(B)** Distribution of the number of routes per target. Comparison between CbyD, USPTO-190,[10] and PaRoutes[8]: **(C)** Distribution of overall complexity,[19] molecular weight, and number of tetrahedral stereocenters of the final target molecules. **(D)** Distribution of the number of aromatic rings, heteroaromatic rings, non-

aromatic heterocycles, and non-aromatic carbocycles per target. **(E)** Percentage of targets containing at least one fused, bridged, or spiro ring system. **(F)** Distribution of the number of O, N, halogen (F, Cl, Br, I), S, and P atoms per target.

Additionally, we compared the molecular complexity of CbyD targets with those of the route datasets commonly used to benchmark multistep CASP, including USPTO-190[10] and PaRoutes.[8] Both are reconstructed from USPTO single-step reaction data.[2] USPTO-190 contains 190 targets with the shortest routes to commercially available starting materials assembled across patents, while PaRoutes contains 10,000 routes reconstructed within individual patents. Both are dominated by medicinal chemistry targets and differ in scope from CbyD.

Because individual complexity metrics quantify different aspects of molecular complexity, we use an overall product complexity score[19] defined as the average of the normalized Böttcher's score,[20] Spacial score (SPS),[21] SAScore,[22] and NPScore.[23] By this measure, CbyD targets show higher molecular complexity than those in USPTO-190 and PaRoutes (**Fig. 3C and Tables S3, S4**). The higher complexity of CbyD targets is not primarily explained by molecular size, as the molecular-weight distributions of CbyD, USPTO-190, and PaRoutes largely overlap. Instead, CbyD targets tend to contain structural features associated with higher molecular complexity, scoring higher on metrics that capture features including non-aromatic rings, $sp^3$ carbons, and stereogenic centers.

CbyD targets also differ from USPTO-190 and PaRoutes targets in their ring systems. CbyD targets are more often polycyclic: about three-quarters contain a fused ring system, compared with 54% in USPTO-190 and 43% in PaRoutes, and about one-third contain a bridged ring system, which is nearly absent from USPTO-190 and PaRoutes (**Fig. 3E and Table S5**). CbyD targets are enriched in non-aromatic rings, with a substantially higher fraction containing three or more non-aromatic heterocycles or carbocycles, whereas USPTO-190 and PaRoutes rarely contain more than two (**Fig. 3D and Table S6**). The opposite trend is observed for aromatic and heteroaromatic rings, which are more prevalent in USPTO-190 and PaRoutes. We catalogued the nitrogen- and oxygen-containing heterocycles present in the CbyD targets (**Fig. S10**). The most frequent rings include tetrahydropyran, tetrahydrofuran, pyridine, piperidine, and indole, consistent with the heterocycles frequently reported in FDA-approved drugs.[24–26] Heteroatom composition shows a similar distinction: nitrogen and halogens are less common in CbyD targets (58% and 13% contain at least one, compared with 88–92% and 48–57% in USPTO-190 and PaRoutes), as is sulfur (7.6% versus 20–26%), while phosphorus is rare in all three datasets (<1%). Oxygen is present in more than 90% of targets across all three datasets, but CbyD contains more oxygen-rich structures, with 27% of targets having six or more oxygen atoms, compared with 20% in USPTO-190 and 7% in PaRoutes (**Fig. 3F and Tables S7, S8**). These differences are consistent with the greater representation of complex natural products in CbyD compared to the patent-derived benchmarks.

**CbyD reactions are more complex than those in patent-derived reaction datasets**

The higher molecular complexity of CbyD targets is also reflected in the reactions used to construct them. We classified reactions in CbyD using NameRxn,[27] a three-level hierarchical reaction classification system, and compared their distributions with those of two large one-step reaction datasets, USPTO-Full[2] and Pistachio.[3] Both datasets are substantially larger than CbyD but rely exclusively on patent data: USPTO-Full is an open-access dataset extracted from U.S. patents, whereas Pistachio is a commercial database of reactions extracted from U.S. and European patents. We observed that CbyD contains fewer heteroatom alkylation/arylation and acylation reactions, which are the two most common classes in USPTO-Full and Pistachio, but a higher proportion of C–C bond formation, protecting-group installation, and oxidation/reduction reactions. CbyD also contains a higher proportion of reactions not assigned to a standard NameRxn class, with 23% classified as *Unrecognized*, compared with 18% in USPTO-Full and 14% in Pistachio (**Fig. 4A and Table S9**). This difference likely reflects both chemistry and classification. Not only has the tradition of "naming" reactions been dwindling since the early 2000s, but modern total syntheses are also likely to contain new reactions and cascade transformations that fall outside the NameRxn rule set. This is unsurprising as academic synthetic practitioners aim to push the boundaries of synthetic chemistry, whereas patented industrially focused chemistry would be expected to favor reliable, well-established transformations that fall into well-defined classes. In addition, only CbyD and Pistachio contain mixed classes, with mixed first-level classes typically involving a main transformation followed by a protection or deprotection step, while mixed second-level classes typically correspond to deprotection of two different types of protecting groups (**Fig. S8**).

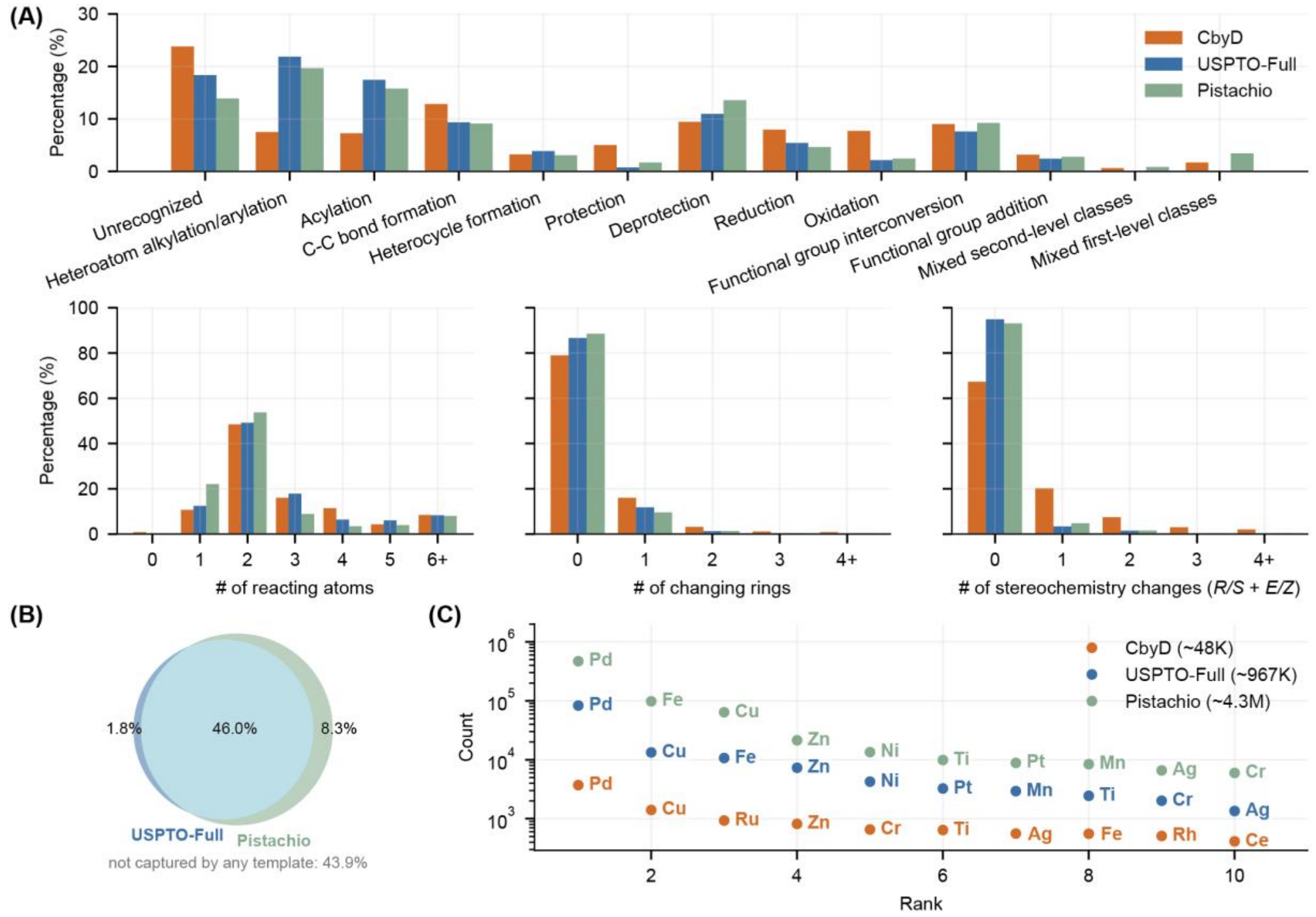

**Figure 4. (A)** Distribution of reaction classes, number of reacting atoms, number of changing rings, stereochemical (*R*/*S* and *E*/*Z*) change. **(B)** Venn diagram reaction-template overlap: Percentage of reactions in CbyD captured by templates extracted from USPTO-Full and Pistachio. **(C)** Rank-size distribution of transition metal usage across CbyD, USPTO-Full, and Pistachio (2025Q2).

Besides differences in reaction classes, we also examined reaction complexity based on the number of reacting atoms and changing rings.[19] A reacting atom is defined as a product atom that exhibits a change in formal charge, neighboring atoms, or bond types relative to its corresponding reactant atom. Changing rings are identified by comparing atom-mapped rings in reactants and products, capturing ring formation, cleavage, and rearrangement. We found that CbyD reactions also involve more reacting atoms and changing rings, and change stereochemistry more often. About one-third of CbyD reactions have at least one stereochemical change, compared with 5–6% in USPTO-Full and Pistachio (**Tables S10, S11**). The products of CbyD reactions are also more complex (**Fig. S7**). These differences raise the question of whether CbyD transformations are represented in patent data at all. To test this, we extracted retrosynthetic reaction templates from USPTO-Full and Pistachio and considered a CbyD reaction covered if at least one template, applied to its product, reproduces the reported reactants (**Section S2.4**).[19] By this criterion, 43.9% of CbyD reactions are not covered by any template extracted from USPTO-Full or Pistachio at RDChiral's default level of template specificity (**Fig. 4B**). Metals used in CbyD also show a distinct trend relative to the patent-derived datasets (**Fig. 4C and Table S12**). Palladium is the most frequently used metal in all three datasets, consistent with the widespread use of palladium-catalyzed cross-coupling reactions in synthetic chemistry.[28] Beyond palladium, frequently used metals in CbyD include ruthenium, rhodium, and cerium, none of which appear among the top ten in USPTO-Full or Pistachio. Ruthenium and rhodium are associated with olefin metathesis and C–H functionalization, respectively, both of which are widely used in complex molecule synthesis, but less common in patent-derived datasets.[29,30]

**CbyD routes are longer, more convergent, and correspond to true experimental precedent**

To assess how CbyD routes differ in length and topology from those in USPTO-190 and PaRoutes, we compared the total number of reactions (TNS) and the longest linear sequence (LLS) across the three datasets.[8,10] As shown in **Fig. 5** and **Table S13**, CbyD routes are generally longer and more convergent than those in the two patent-derived benchmarks, with a median LLS of 13 steps compared with 6 and 3 for USPTO-190 and PaRoutes, respectively; routes on the diagonal are linear, whereas those above the diagonal are convergent. Similarly, 38.6% of CbyD routes are convergent, compared with 28.9% and 3.0% in the other two datasets. Whereas PaRoutes is capped at a depth of 10 by construction, the longest CbyD route along the LLS is Isobe's linear 73-step synthesis of tetrodotoxin. In contrast, Danishefsky's rapamycin synthesis has the largest TNS of 114, but an LLS of only 45, because most of its steps prepare fragments that are later combined.[31,32] These longer and more convergent routes reflect the complexity of CbyD targets, which could serve as challenging test cases for multistep CASP methods.

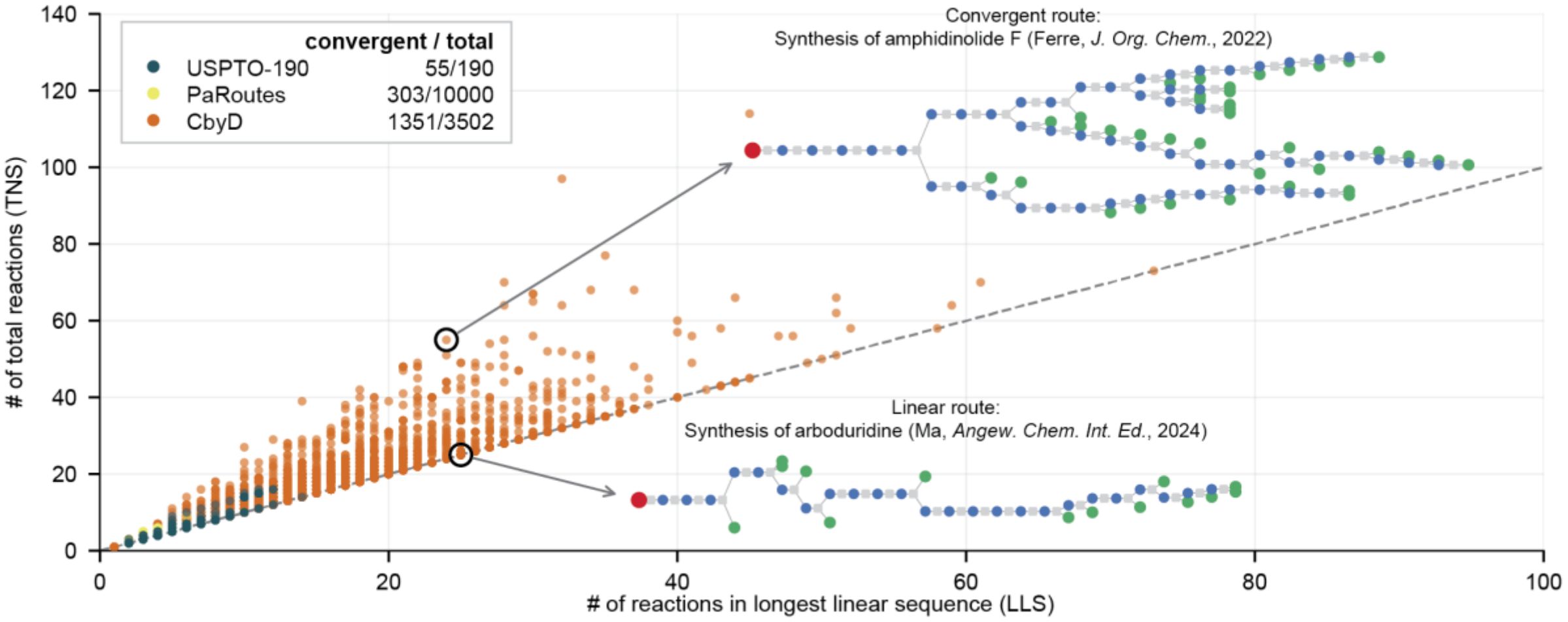

**Figure 5.** Number of total reactions (TNS) versus number of reactions in the longest linear sequence (LLS) for CbyD (orange), USPTO-190 (teal) and PaRoutes (yellow). Points on the dashed diagonal are linear routes; points above it are convergent. The insets show one linear and one convergent CbyD route, in which the target is red, intermediates blue and starting materials green.

These differences also reflect how the datasets are constructed. USPTO-190 and PaRoutes are reconstructed from one-step reaction data by connecting reactions where the product of one reaction serves as a reactant in another, then extracting synthetic routes from the resulting reaction networks. As a result, the reconstructed routes may not correspond to complete synthetic sequences that have been experimentally executed (**Section S4.1**). These artificially reconstructed routes show pathologies that would confuse models if used for training, and lead to erroneous conclusions if used for evaluation. In contrast, CbyD records complete routes reported in the literature following peer review by expert synthetic chemists.

### Starting materials vary in commercial accessibility and contribution to stereochemical complexity

How much complexity a route must build depends not only on its transformations but also on the starting materials from which it begins. Chiral starting materials are of particular interest because they can introduce the stereochemistry of the target molecule, either through the incorporation of existing stereocenters or by directing the formation of new ones. Frequently used chiral starting materials in CbyD fall into two classes (**Fig. 6**). The first is the classical chiral pool, including terpenes such as carvone and limonene, $\alpha$-amino acids and their esters, hydroxy acids such as the Roche ester, malic acid, and ethyl lactate, and small chiral epoxides such as glycidol, epichlorohydrin, and propylene oxide.[33,34] These chiral starting materials provide stereocenters that can be carried directly into the target structure. The second consists of chiral auxiliaries, including Evan's oxazolidinones, Ellman's *tert*-butanesulfinamide, and Myers pseudoephedrine amides.[35–37] These reagents do not contribute atoms to the final product but impart stereochemical information during a reaction and are subsequently removed.

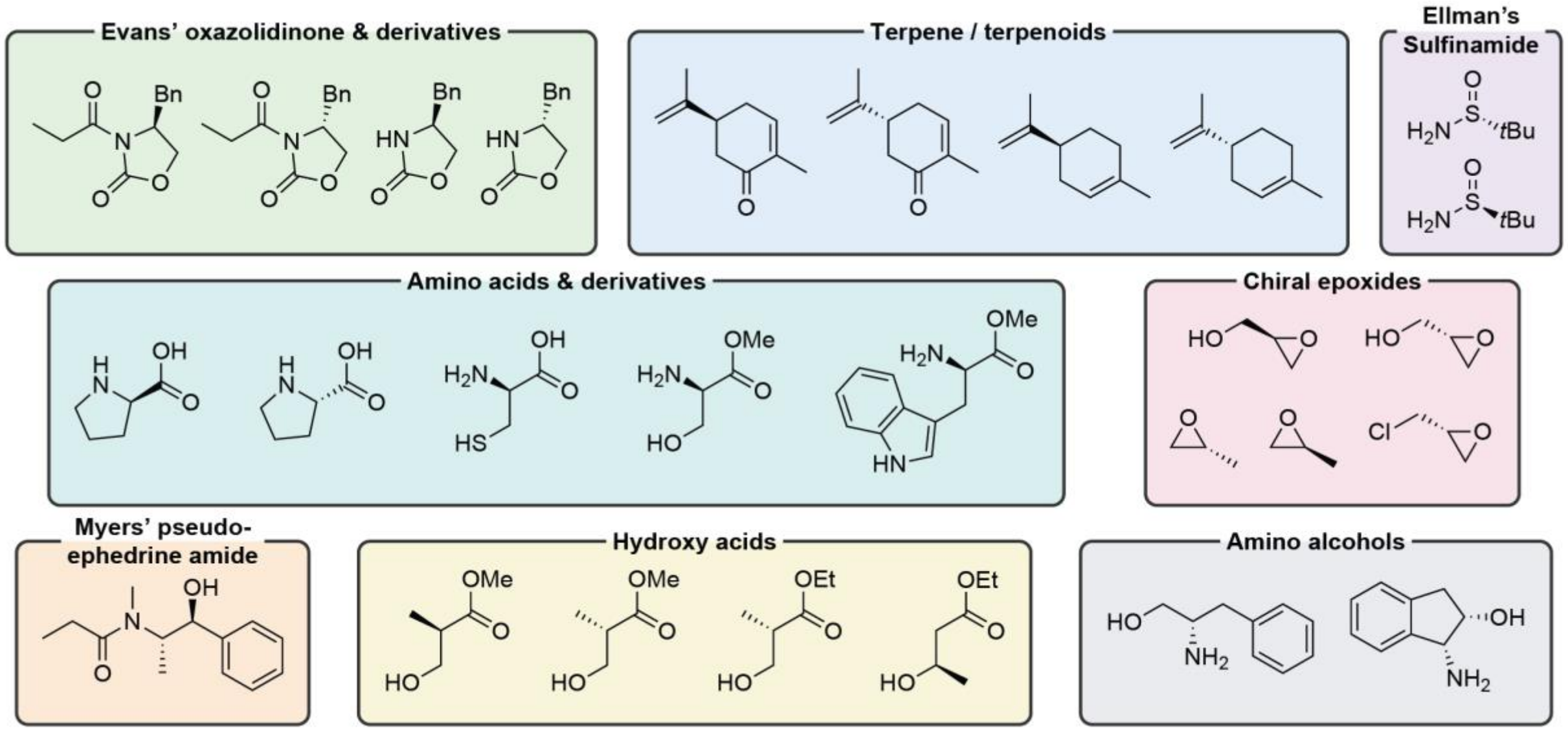

**Figure 6.** Representative frequently used chiral starting material classes in CbyD routes.

To quantify the contribution of chiral starting materials, we traced each target stereocenter back through the route using atom-mapping to determine whether it was inherited from a starting material or formed during the synthesis (one example is shown in **Fig. S13**). Only 17% of target stereocenters are directly inherited from chiral starting materials; the remaining 83% are formed during the route. Therefore, most of the stereochemistry in CbyD targets is set by stereoselective reactions or chiral resolution within the route. The 17% therefore represents a lower bound on the contribution of chiral starting materials to target stereochemistry, as chiral auxiliaries can also control the formation of stereocenters without being incorporated into the final target structure.

Beyond their role in controlling stereochemistry, the availability of starting materials is also important when considering CbyD as a potential CASP benchmark. In curating CbyD, we traced the original publications, its accompanying supporting information, and used cited references or performed separate database searches to identify commercially available starting materials whenever possible. This was a critical design decision when the website and app were first created, wherein the goal was to capture the totality of reactions employed to complete a given target molecule, instead of focusing solely on reactions associated with LLS, from sources that would have been deemed reasonably priced as starting materials for academic and industrial applications. However, CASP methods rely on predefined purchasable-compound databases, which do not necessarily cover all starting materials used in published syntheses. For example, we found that 29.0% of unique CbyD starting materials are absent from one exemplary building block library from eMolecules (similar to what has been used in prior retrosynthesis planning efforts),[38] including many common reagents that it does not seek to cover (**Section S3.3**). Routes containing these starting materials or reagents would not be able to be recovered if using only this database. It is important to consider starting-material database coverage and accessibility when evaluating CASP methods.

**Molecular complexity rises sharply at a small number of key transformations**

While the complexity of starting materials provides information about the initial stage of a synthesis, changes in molecular complexity along the route can help identify transformations and synthetic blueprint design choices such as the union of key fragments that contribute substantially to the complexity of the target. A notable exception to this is the addition of protecting groups which increase structural complexity without adding value in terms of new bonds or atoms found in the final target structure to the overall sequence at that particular step. We examined these changes along the longest linear sequence of each route using four metrics: Böttcher's score,[20] SAScore,[22] SPS,[21] and NPScore.[23] Böttcher's score, SAScore, and SPS generally increase toward the target (**Figs. 7A**, **S14**), whereas NPScore remains nearly flat over the first half of the route and varies substantially across routes. This behavior is consistent with NPScore measuring natural-product-likeness rather than structural complexity. Thus, the choice of metric also determines which steps appear to drive the largest increase in complexity (**Fig. S14**).

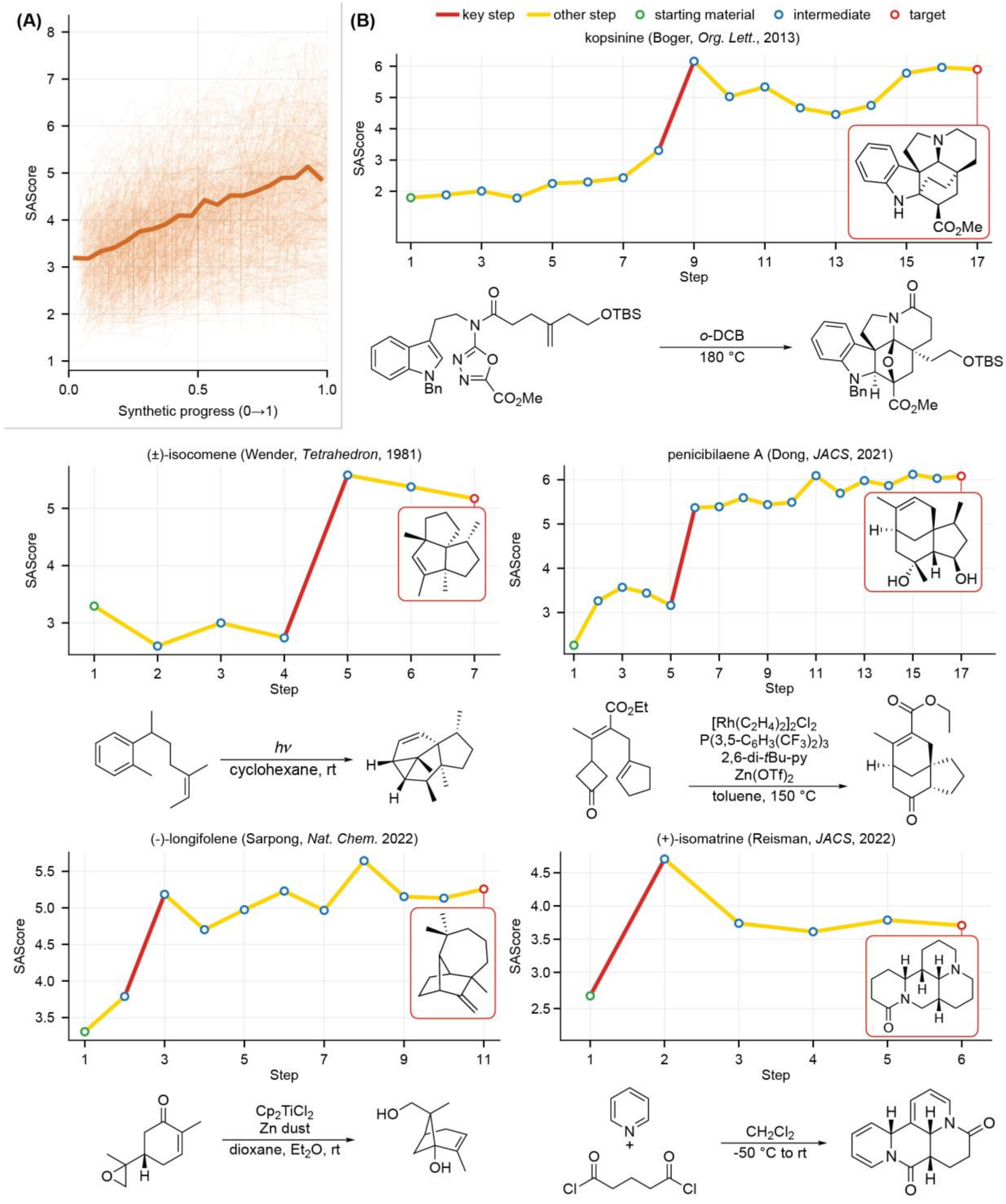

**Figure 7. (A)** Molecular complexity SAScore along CbyD routes, plotted against normalized synthetic progress, defined as the step number divided by the total number of reactions along the longest linear sequence (LLS). Thin lines show individual routes, and the thick line shows the mean trend. **(B)** SAScore of each intermediate across the synthesis of five representative targets. Green and red markers denote the starting material and target, respectively. Red segments mark sharp increases in SAScore; the transformation corresponding to each is shown below. Full route trees are provided in **Section S3.4**.

Within individual routes, complexity does not increase uniformly. Most routes show relatively small SAScore changes over several consecutive steps, with one or two sharp increases that often coincide with core-skeleton formation. Complexity can also decrease in subsequent steps as protecting groups or chiral auxiliaries are removed and functional groups are modified. The five examples in **Fig. 7B** illustrate how different types of ring-forming transformations can produce these sharp increases in complexity. In the syntheses of isocomene and kopsinine, cycloadditions expediently construct polycyclic skeletons: an intramolecular arene–alkene meta-photocycloaddition forms the tetracyclic intermediate of isocomene, while a cycloaddition cascade forms the pentacyclic skeleton of kopsinine and establishes five contiguous stereocenters. In both cases, the groups leading these synthetic efforts recognized the immense untapped potential of these new transformations and strategies and decided to showcase in natural product total syntheses.[39,40]

Other examples involve distinct ring-forming strategies, including a dearomative pyridine annulation that constructs the tetracyclic core of isomatrine resulting in a remarkable 5-step total synthesis from readily available building blocks[41], a Rh-catalyzed C–C activation that assembles the all-carbon quaternary center and the bridged tricyclic core of the penicibilaenes[42], and a titanium-mediated radical cyclization that generates the bicyclo[3.1.1]heptanol framework of longifolene, a natural product that has been a playground of synthetic chemists since the 1960's, in the shortest LLS reported to date.[43] Across these routes, SAScore changes by less than one unit per step for most of the sequence but increases by approximately 1–3 units at these key transformations. These examples illustrate how a single transformation can account for a substantial increase in molecular complexity, particularly through the construction of complex ring systems. Such transformations, including cycloadditions, dearomatizations, and other ring-forming reactions, are less frequently represented in patent-derived datasets, consistent with the reaction class distribution discussed earlier (**Fig. 4A**).

## CONCLUSION AND OUTLOOK

Since its launch in 2011, CbyD has served as an open-access educational resource for learning organic reactions and synthetic strategies through published total syntheses of natural products and pharmaceuticals. In this work, we present the first release of CbyD as a structured, machine-readable dataset for reaction informatics and computer-aided synthesis planning. The dataset provides a consistent representation of published synthetic routes, including individual reactions, reagent information, and literature sources.

We characterized CbyD at both the reaction and route levels, comparing its composition with existing reaction and route datasets. CbyD contains more complex targets and reactions, as well as longer and more convergent routes. Most target stereocenters are set by reactions within the route rather than inherited from chiral starting materials, and molecular complexity typically increases in one or two sharp steps that coincide with core-skeleton formation.

These features provide opportunities for developing and evaluating CASP methods at multiple levels. At the reaction level, CbyD contains complex transformations that are sparsely represented in patent-derived datasets, providing a more challenging setting for evaluating single-step retrosynthesis models. At the route level, its longer and more convergent syntheses allow methods to be evaluated on whether they can identify productive intermediates and combine independently prepared fragments over long sequences. Because each record contains a complete published route, recorded intermediates can also serve as planning constraints with known solutions, such as completing a synthesis from an advanced intermediate or finding a route through a specified intermediate.[44,45]

Complex targets remain one of the most challenging settings for synthesis planning, where successful routes depend on decisions made across the entire route: when to install a stereocenter, when to protect or deprotect a group, and how early steps set up the key disconnections that follow.[7] We have outlined several analyses enabled by CbyD, and its potential for being a challenging benchmark for multistep CASP methods, although we do not train or evaluate those in this work. Such evaluations are left for future studies, and the broader value of CbyD will likely extend to questions that we have not yet anticipated. We hope the dataset will provide a basis for computational and synthetic chemists to investigate new questions about route design and evaluate methods against published synthetic practice.

CbyD and its curation process also have some limitations. Reaction conditions are primarily extracted from the ChemDraw schemes, which do not always include the detailed experimental procedures reported in the original literature. As a result, information such as reagent quantities, reaction times, and workup procedures is not available. In addition, the curation process required substantial manual effort, particularly in preparing the ChemDraw routes and reviewing the converted data. Recent advances in vision-language models for chemical structure and reaction recognition[46–49] will help automate parts of this process, allowing researchers to extract synthetic routes directly from published literature and contribute them to CbyD with less manual effort.

CbyD will continue to be expanded and maintained through versioned public releases, so that results reported on a given release remain reproducible as records are added and corrected. We encourage researchers to contribute new routes, report errors in existing records, and add information that may be useful for future analyses. Collecting published syntheses in an open, curated, and machine-readable form is a shared effort, and we hope CbyD can serve as a common foundation for it, bringing CASP closer to the way synthetic chemists design routes.

## ASSOCIATED CONTENT

**Data and standalone dashboard availability.** The CbyD dataset is publicly available on GitHub under the Creative Commons Attribution 4.0 International (CC BY 4.0) license at https://github.com/chemistry-by-design/cbyd.

**Code availability.** The code used for data processing and analysis is available on GitHub under the MIT license at https://github.com/chemistry-by-design/cbyd_analysis.

## AUTHOR INFORMATION

### Corresponding Authors

*Email: ccoley@mit.edu, njardars@arizona.edu

### Author Contributions

†These authors contributed equally.

## ACKNOWLEDGMENTS

JTN thanks the National Science Foundation (CHE-2449261) for their generous support. SBAT, JR, KY, and CWC were supported by the NSF C-CAS under Grant CHE-2202693 and the Machine Learning for Pharmaceutical Discovery and Synthesis consortium. KY thanks NSF

GRFP program for additional funding. JR thanks the XtalPi AI for Science Fellowship and the Takeda Fellowship for additional funding. The authors thank the MIT Office of Research Computing and Data for providing the computing resources that have contributed to the research results reported within this paper. ZJ and PS acknowledge support from the Swiss National Science Foundation (SNSF) [214915]. The authors thank Daniel Armstrong, Théo Neukomm, and Andres M. Bran from the Schwaller group for insightful discussions in the early stage of the project.

# Supporting Information for:

# Chemistry by Design (CbyD): A Machine-Readable Dataset of Expert-Curated Total Synthesis Routes

*Ryan E. Williams[1,†], Suong B. A. Tran[2,†], Hayden M. Leatherwood[1], Jihye Roh[3], Aquila Wolff[3], John G. Federice[1], Nicholas R. Lauta[1], Kevin Yu[4], Steven J. Nick[1], Zlatko Jončev[5], Philippe Schwaller[5], Connor W. Coley[3,4,*], Jon T. Njardarson[1,*]*

[1]Department of Chemistry and Biochemistry, University of Arizona, Tucson, AZ, USA

[2]Department of Chemistry, Massachusetts Institute of Technology, Cambridge, MA, USA

[3]Department of Chemical Engineering, Massachusetts Institute of Technology, Cambridge, MA, USA

[4]Department of Electrical Engineering and Computer Science, Massachusetts Institute of Technology, Cambridge, MA, USA

[5]Institute of Chemical Sciences and Engineering, École Polytechnique Fédérale de Lausanne, Lausanne, Switzerland

[†]These authors contributed equally.

*Email: ccoley@mit.edu, njardars@arizona.edu

## S1 DETAILS ABOUT DATA CURATION PROCESS

The CbyD collection was originally drawn in ChemDraw[1] with each target synthesis saved as individual ChemDraw file before each starting material, product and reagents were converted into images files prior to application and website upload. Converting it into the machine-readable format described in **Fig. 2** required (i) an automated extractor that reads the ChemDraw CDXML files and writes one JSON record per route, and (ii) a manual inspection pass in which extracted routes were checked against the original drawings and corrected.

### S1.1 Extraction pipeline

Each file is processed as follows (**Fig. S1**):

- **Reaction extraction.** Reactions were extracted from ChemDraw files by parsing the CDXML document tree. Starting from the root, each scheme element was traversed to its constituent step elements, each of which encodes a single reaction transformation. A step carries identifier lists for `ReactionStepReactants`, `ReactionStepProducts`, and `ReactionStepObjectsAboveArrow`.
- **Identifier resolution.** The identifiers are resolved against the document: fragment elements become molecules (**Section S1.2**) and text elements above the arrow are collected as condition text (**Section S1.5**).
- **Reaction assembly.** Reactants, products, and any molecules drawn above the arrow are written to `reaction_smiles` in the format `reactants>agents>products`, preserving the information extracted directly from the ChemDraw file.
- **Agent classification and atom mapping.** Dictionary SMILES for the identified reagents, catalysts, ligands, and solvents are merged onto the reactant side and the resulting reaction is atom-mapped using RXNMapper.[2] The mapped reaction SMILES and associated confidence score are stored as `reaction_smiles_mapped` and `mapping_confidence`, respectively, in `reactants>agents>products` format. Here, the reactant field contains molecules that contribute atoms to the final product. We also checked for chemically equivalent reactants.

- **Manual final check.** Each route is checked against the original source to verify full route connectivity, SMILES validity and accuracy, and atom-mapping correctness.

Note: Any file containing an asterisk (*) is rejected since an asterisk indicates an undefined attachment point or R group that cannot be resolved to a structure. Such files must be repaired before extraction is attempted again.

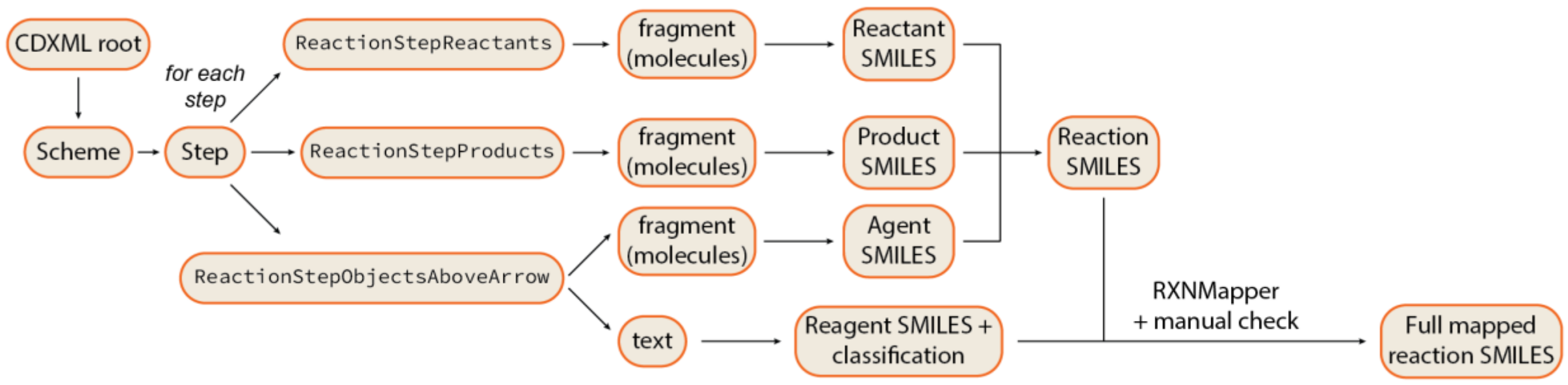

**Figure S1.** Overview of reaction SMILES extraction pipeline

## S1.2 Molecule construction

Each fragment element is converted independently:

- The fragment is wrapped in a minimal

  `<CDXML BondLength="14.4"><page>…</page></CDXML>`

  envelope and passed to `rdkit.Chem.AllChem.MolsFromCDXML`, which builds the molecule from the atomic coordinates and bond display attributes and perceives tetrahedral centers from wedged and hashed bonds.

- Two bond display attributes are rewritten in the fragment XML: `Display="Bold"` becomes `Display="WedgeBegin"`, and `Display="Hash"` becomes `Display="WedgedHashBegin"` (**Section S1.3.1**).
- `SanitizeMol` and `AssignStereochemistry(force=True, cleanIt=True)` are applied, and the molecule is written with `MolToSmiles(mol, isomericSmiles=True).`

## S1.3 Handling stereochemistry

Stereochemistry in the JSON files is determined by RDKit's perception of wedge and hash bonds. RDKit assigns chiral tags based on neighbor ordering, from which the corresponding *R*/*S* descriptors are determined according to CIP priority. Stereochemistry is the largest source of error in the automated extraction and can also occur when using ChemDraw's SMILES generator. During the extraction process we learned that wedges of stereocenters were sometimes read in the wrong direction, causing stereocenters to invert, and many schemes used bold bonds instead of wedged bonds, which ChemDraw doesn't interpret as stereochemistry. These stereochemical errors we encountered were often limited to a few intermediates, leading to a route beginning and ending correctly still having a few incorrect structures along the way. This section describes how we handled these types of stereochemistry related challenges.

### *S1.3.1 Bold/Hash bonds*

In ChemDraw, `WedgeBegin` and `WedgedHashBegin` are stereo bonds with a defined narrow end, from which ChemDraw computes *R*/*S*. Bold and Hash are display-only styles that resemble a wedge or a hash but have uniform width. ChemDraw does not treat a bold bond as a stereo bond when generating SMILES, so a center who's only stereochemical cue is a bold bond is omitted without a chirality marker. To recover these lost stereocenters, the extractor rewrites `Bold` to `WedgeBegin` and `Hash` to `WedgedHashBegin` before RDKit reads the XML, so that RDKit interprets the bond as a wedge originating at the bond's begin atom (**Fig. S2A**). This rewrite is only a partial remedy. A bold bond has no intrinsic narrow end, so the substitution assigns configuration at the bond's begin atom alone. Where a bold bond joins two stereogenic ring atoms, the second center is defined only if another wedge or hash elsewhere in the drawing constrains it. Bold bonds between two stereocenters were consequently a recurring source of incorrect or missing stereochemistry and were resolved by manual inspection and redrawing.

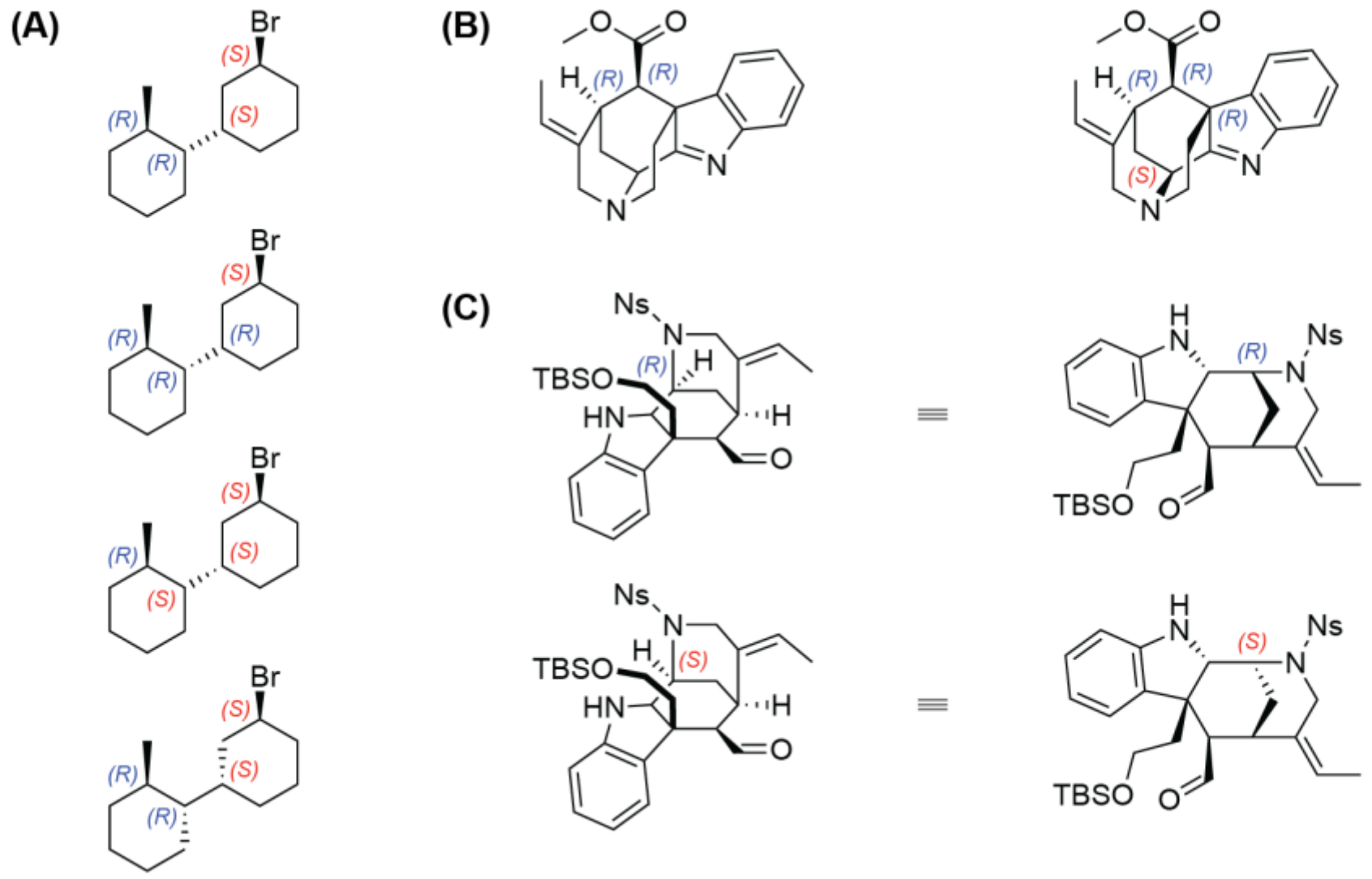

**Figure S2.** Common sources of stereochemical ambiguity in drawn structures. **(A)** Bold bonds used to convey perspective are interpreted as stereo wedges, so small changes in where the bold bond is drawn alter the assigned CIP descriptors of otherwise identical structures. **(B)** Bridgehead stereocenters that are implied by the ring system but not explicitly defined must be added to fully specify the molecule; left, as drawn; right, with the bridgehead stereocenter assigned. **(C)** The same caged scaffold drawn in perspective (left) and redrawn in a flat representation (right); a slight change in how the hydrogen is drawn inverts the configuration at that stereocenter.

### *S1.3.2 Bridgehead centers and perspective drawings*

Bridgehead stereocenters is another case in which the drawing style, rather than the extractor, determines the outcome. The affected structures are almost all drawn in a three-dimensional or perspective style, with the ring system shown as a cage and no wedges or hashes on the bridgehead bonds. Older versions of ChemDraw cannot perceive stereochemistry from that geometry, and their SMILES output carries no chirality markers at those centers. Where the released JSON files do contain bridgehead stereochemistry, it is because wedges and hashes were drawn explicitly onto the bridgehead bonds in the ChemDraw file before extraction, so that both ChemDraw and RDKit had unambiguous stereo bonds to work from (**Fig. S2B**). Even then the result depended on how the ring system was drawn: the same bridgehead was assigned $R$ or $S$ depending on the perspective of the drawing (**Fig. S2C**), so these centers were verified by hand and redrawn where necessary.

The RDKit version used for extraction therefore matters. Many JSON files were generated with RDKit releases from 2024 or earlier, which do not perceive bridgehead stereochemistry from ring geometry alone and may assign no chirality to an unassigned bridged bicyclic structure. RDKit 2025.03 and later can perceive bridgehead stereochemistry from 2D coordinates, allowing these stereocenters to be recovered during re-extraction without manually drawn wedges. To ensure accurate stereochemical assignments, we identify affected structures and re-extract and manually inspect them for correction.

## S1.4 ChemDraw behaviors requiring special handling

### *S1.4.1 Contracted abbreviation labels*

Contracted labels such as -Boc, -NHBoc, and -OTBS are represented as nested fragments with attachment-point metadata. The extractor passes the complete fragment element to

`MolsFromCDXML`, preserving the attachment information and correctly reconstructing the molecular structure. However, labels that are not defined in ChemDraw may fail to expand during extraction (**Fig. S3A**). The way these labels are written can also affect the resulting SMILES (**Fig. S3B**). These cases therefore require careful inspection after extraction.

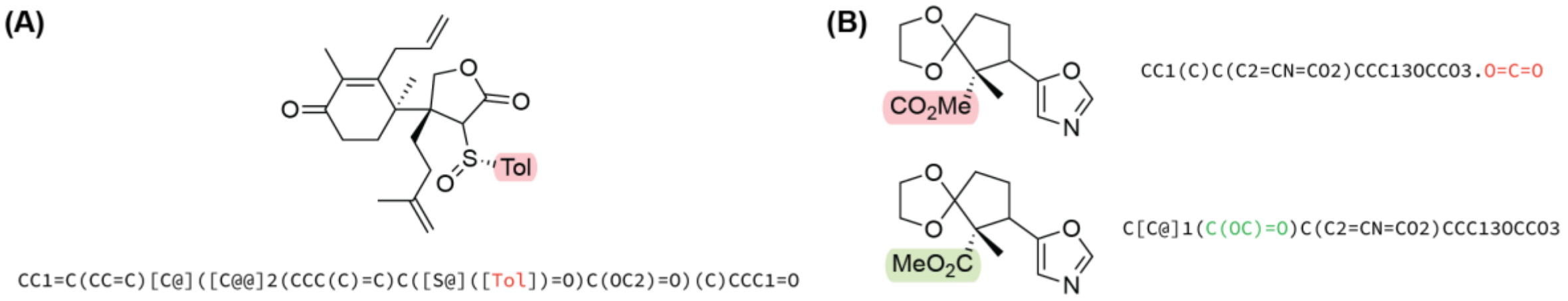

**Figure S3.** Errors introduced by text labels and abbreviations during structure-to-SMILES conversion. **(A)** A sulfinyl substituent whose label is partially condensed in the drawing is exported with an unexpanded abbreviation, giving a SMILES string containing the unresolved pseudo-atom `[Tol]`. **(B)** The same ester drawn with two label orientations. Top: is misparsed as a disconnected fragment plus a methyl group, and the stereocenter is lost. Bottom: is correctly interpreted as a methyl ester with the stereocenter retained.

### *S1.4.2 Name-to-structure misinterpretation*

When a reagent name in a condition text box is set to “interpret as chemical”, ChemDraw parses the abbreviation letter by letter into a chemically meaningless structure (**Fig. S4**). This was the most pervasive failure mode encountered in the extraction, alongside the stereochemical issues described above. Reagent identities in the JSON files are therefore taken from curated dictionaries keyed on the text label (**Section S1.5**), and never from ChemDraw’s interpretation of that label.

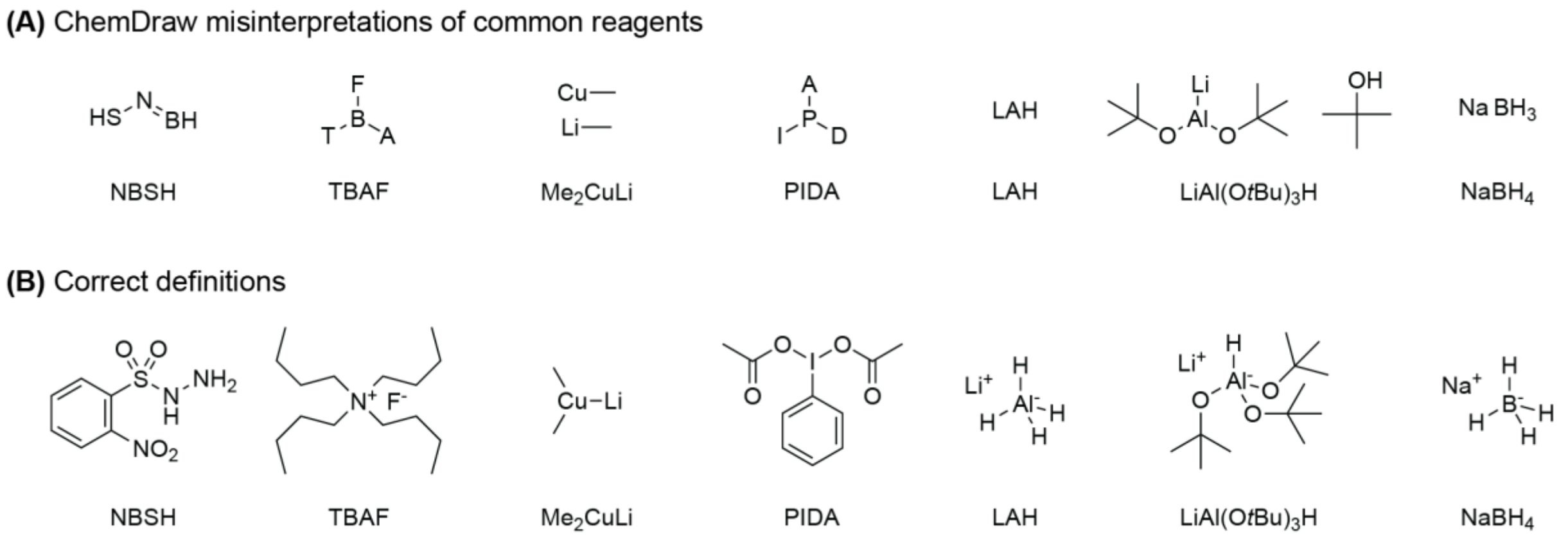

**Figure S4.** ChemDraw's interpretation of common reagent abbreviations compared with the curated dictionary definitions.

### *S1.4.3 Miscellaneous*

- **Objects below the arrow:** Only `ReactionStepObjectsAboveArrow` is read; anything placed below the arrow is not captured. This is by design: in CbyD all reaction conditions are placed above the arrow, and files were reformatted to that convention before extraction.
- **Character encoding:** Files saved on different systems produced mis-decoded degree symbols, in which the degree sign is preceded by a stray Â character. The condition normalizer removes the stray character, unifies `° C, deg C, and degC` to a single `°C` form, and strips the word "then" from tokens.

## S1.5 Classification of condition text

Each token obtained from the condition text is processed using a set of curated JSON dictionaries: `CATALYSTS, LIGANDS, SOLVENTS, MODIFIERS`, `GENERAL`, `TEMPERATURE`, `YIELD`, and `IGNORE`. Matching is literal, so dictionary keys must correspond exactly to the way chemical species are written in the drawings. Each token is assigned to exactly one category according to the following precedence: `GENERAL_REAGENTS > CATALYSTS > LIGANDS > SOLVENTS > MODIFIERS`. Tokens not matched to a more specific category are recorded as general reagents. `IGNORE.json` contains terms that are excluded from reagent mapping, including reaction names, placeholders, stoichiometric or condition fragments, and generic labels that do not identify specific chemical species. `MODIFIERS.json` contains reaction conditions, media, physical setups, resins, and biocatalysts that are recorded as modifiers rather than stoichiometric reagents.

For identified chemical species, the corresponding dictionary SMILES are stored in the …`_DEFINITIONS` fields and used to construct reaction SMILES that is then atom-mapped. Temperature and yield are processed separately. Yield entries are cleaned to retain the numerical yield (%), yield qualifiers such as the number of steps and BORSM, dr/ee/er ratios, and "quant."; all other text is removed. For example, "$CH_2Cl_2$, rt, 88%" is reduced to "88%", while "TfOH, THF, 0 °C, 75% (2 steps)" is reduced to "75% (2 steps)". Retained yields are then reformatted into the canonical form xx% (qualifier), with qualifiers for steps, dr, er, ee, *E*/*Z*, and BORSM placed in parentheses. For example, "55%, 2 steps" becomes "55% (2 steps)", "98 : 2 d.r., 80%" becomes "80% (dr=98:2)", and "80%, 93% ee" becomes "80% (ee=93%)".

### S1.6 Drawing conventions for future contributions

The failure modes described above reflect properties of the ChemDraw files as much as limitations of the extractor and were identified during the construction of the dataset. Moving forward, to facilitate future expansion of the dataset, the following conventions were adopted for CbyD files and are recommended for contributed files:

- Denote stereochemistry with directional wedged and hashed bonds only. Bold and plain hash display styles must not be used to indicate configuration (**Section S1.3.1**).

- Draw complex ring systems, such as sugars, steroids, alkaloids, and complex terpenoids as a flat, skeletal structure with stereocenters unambiguously assigned with directional wedged and hashed bonds. Avoid 3-dimensional conformational projections such as chairs and boats (**Section S1.3.1**).
- Assign bridgehead and other cage stereocenters explicitly with wedges or hashes rather than relying on a perspective drawing (**Section S1.3.2**).
- Leave reagent and condition labels as plain text; do not set condition text to be interpreted as a chemical structure, which is often the default (**Section S1.4.2**).
- Place all reaction conditions, reagents, and yields above the arrow.
- Avoid undefined attachment points and R groups; draw structures explicitly.
- Use abbreviation labels that ChemDraw can expand and avoid reagent names containing commas in condition text.

### S1.7 Note on tree route construction

Construction of the synthesis tree for CbyD requires additional considerations arising from the connectivity and structural roles of molecules across individual reactions. The following rules were applied during tree construction:

- Each molecule is assigned a structural role as a target, intermediate, starting material, or byproduct based on whether it is consumed or produced within the route.
- Molecules appearing as reagents are tracked throughout the route, allowing unconsumed reagent-derived molecules to be distinguished from the target and assigned as byproducts.
- No reaction is silently removed during tree construction. Reactions that cannot be parsed are retained as invalid, reactions in which the reactants and products are unchanged, such as salts or complexes, are retained with a warning, and chemically valid reactions that cannot be connected to the remainder of the route are retained as orphan reactions.

Therefore, in addition to the reaction and metadata representation described above, we release the tree representation of each CbyD route, providing a structured representation of the connectivity and molecular roles within the complete synthesis.

```
{
  "target": "<SMILES>",      % root molecule node (final target)
  "type": "mol",
  "children": [
    {
    "type": "reaction",
    "reaction_id": "<id>",
    "reaction_smiles_mapped": "<reactants>agents>product>",
     "children": [
       {                       % one molecule node per reactant
        "smiles": "<SMILES>",
        "type": "mol",
        "children": [ ... ]    % starting-material leaves
        }
      ]
    }
  ]
}
```

## S2 METHOD

### S2.1 Tools version

All analyses were performed in Python 3.10 with RDKit 2024.09.4 (unless otherwise specified in the corresponding section),[3] NumPy 1.26.4, pandas 2.1.4, and NetworkX 3.4.2 for route graphs and matplotlib 3.10.5 for plotting. The Böttcher score was computed with Open Babel 3.1.1 and the molcomplex code from ASKCOS (https://gitlab.com/mlpds_mit/askcosv2/molecular_complexity).

## S2.2 Datasets

CbyD contains 3,502 literature total-synthesis routes comprising 56,640 reaction steps. Route-level comparisons used PaRoutes set n1 (10,000 routes)[4] and USPTO-190 (190 routes).[5] Reaction-level comparisons used USPTO-Full[6] and Pistachio 2025Q2 (NextMove Software)[7] after cleaning, which retained 966,744 of 1,104,587 USPTO-Full reactions and 4,326,750 Pistachio reactions. Cleaning removed duplicate and malformed reaction SMILES, reactions labeled only as NameRxn class 11, products with fewer than five heavy atoms and multi-product reactions; molecules without mapped atoms were reassigned to agents, the original atom maps were retained, and reactions were deduplicated on their map-free reactants and product. For the reaction-level comparisons, CbyD reactions were deduplicated in the same way, keeping the first occurrence in route order, which left 48,155 unique reactions.

## S2.3 Descriptors

### *S2.3.1 Molecular descriptors*

Molecular descriptors were computed once per unique target with RDKit 2024.09.4, except the Böttcher score. SAScore,[8] which ranges from 1 (easy) to 10 (difficult to synthesize), and NPScore,[9] a measure of natural-product likeness, were computed with the RDKit Contrib implementations, and SpacialScore (SPS)[10] with RDKit without size normalization. The Böttcher score[11] was computed with Open Babel 3.1.1[12] through the molcomplex implementation from ASKCOS[13], in turn adapted from Wright et al.,[14] using the stereocenters of the input SMILES and a correction for symmetry-equivalent atoms. Overall target complexity is the unweighted mean of these four scores after min–max normalization of each over the pooled targets of CbyD, USPTO-190 and PaRoutes. Molecular weight is the RDKit average molecular weight, and tetrahedral stereocenters were counted with `FindMolChiralCenters`, including unassigned centers. Rings were taken from the RDKit symmetrized smallest set of smallest rings and classed as aromatic (including heteroaromatic), heteroaromatic, non-aromatic heterocyclic or non-aromatic carbocyclic. Two rings that share exactly one bond form a *fused* junction and two rings that share two or more bonds a *bridged* junction; an atom shared by two ring systems, rings joined through shared bonds forming one system, is a *spiro* junction. Heteroatoms were counted as O, N, halogens (F, Cl, Br and I), S and P.

### *S2.3.2 Reaction descriptors*

Reaction descriptors were computed with RDKit 2024.09.4 from the atom maps (RXNMapper 0.4.2 for CbyD, the source maps for USPTO-Full and Pistachio). Reacting atoms and changing rings were counted as described previously[15]; all 48,155 unique CbyD reactions could be parsed. C–C and C–X bonds formed are bonds between mapped atoms that are present in the product but not in the reactants, X being any heavy atom other than carbon. Stereochemical changes were assessed by configuration rather than by CIP label, because $R/S$ and $E/Z$ labels can change with substituent priorities alone. On map-free copies of both sides, a mapped tetrahedral center counts as changed if it is created, destroyed or inverted (chiral tag combined with the permutation parity of its mapped neighbors) or if a bond to it is formed or broken, and a double bond present on both sides counts as changed if its configuration appears, disappears or inverts relative to fixed mapped reference substituents.

CbyD reactions were classified with NameRxn 3.7.7 (NextMove Software)[16] (options `-ismi -ordf -matching -addrxnname`) on map-free reaction SMILES, keeping all assignments; a reaction with several assignments was labeled *mixed sub-classes* when they share one superclass and *mixed classes* otherwise, and unrecognized reactions were assigned class 0. USPTO-Full and Pistachio labels were taken from the source data NameRxn 3.7.7. Transition metals (d-block elements of periods 4–6 and the lanthanides) were counted as metal-containing fragments in the agents; metals mapped into the product are reactants and were not counted.

### S2.4 Template extraction

Template coverage was evaluated with the template-enumeration workflow described previously.[15] Retrosynthetic templates were extracted from the atom-mapped USPTO-Full and Pistachio reactions with the RDChiral template extractor[17] as implemented in ASKCOS.[13] Each template includes the reaction center (mapped atoms whose connectivity or properties change) expanded to its first neighbors (radius 1), together with predefined special functional groups adjacent to the reaction center (e.g., carbonyls, alkenes, and common protecting and leaving groups) and all unmapped reactant atoms. Each template was applied to the product of every CbyD reaction with RDChiral using RDKit 2020.09.5**,** consistent with the version used for template extraction, and a CbyD reaction was considered covered if at least one template reproduced its recorded reactants.

## S3 ADDITIONAL RESULTS

### S3.1 Target complexity

The comparison of target complexity in the main text is based on an overall complexity score defined as the average of four normalized metrics.[15] Because each of these metrics captures a different aspect of molecular complexity, and because an aggregate score can obscure disagreement between its components, we report the underlying distributions individually here (**Fig. S5**).

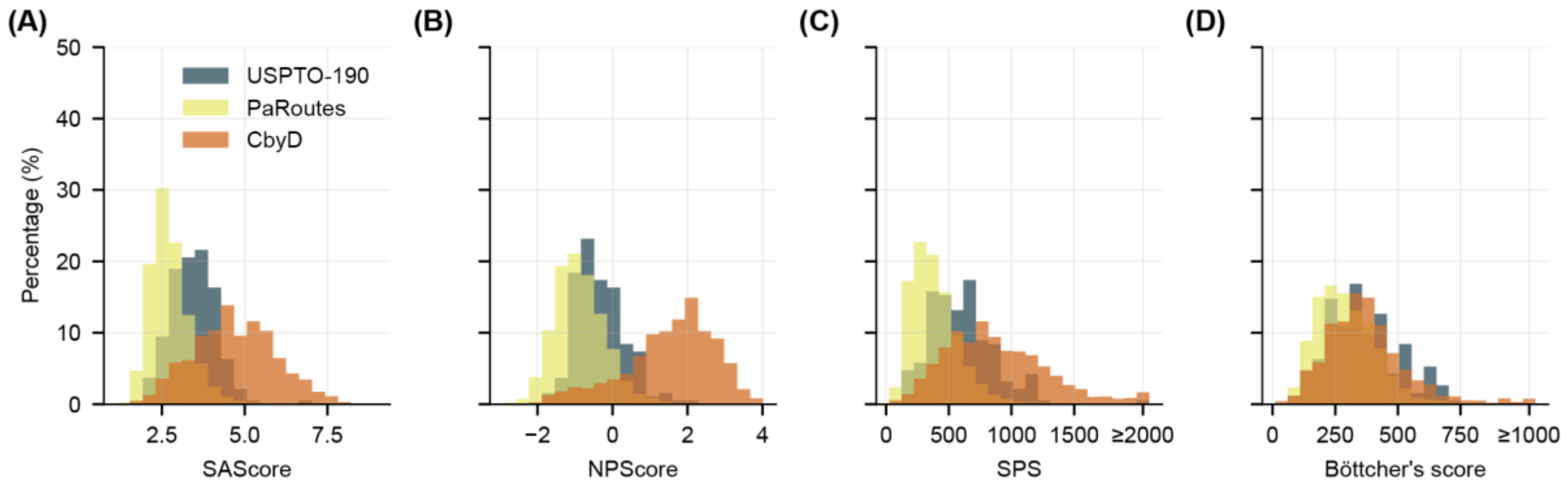

**Figure S5.** Distribution of molecular complexity metrics: **(A)** SAScore, **(B)** NPScore, **(C)** SPS, **(D)** Böttcher's score for target molecules in CbyD, USPTO-190, and PaRoutes.

The four metrics are complementary rather than interchangeable. SAScore[8] estimates synthetic accessibility from the frequency of a molecule's fragments in known compounds together with penalties for size, rings, and stereocenters. NPScore[9] measures how closely the fragment composition of a molecule resembles that of natural products rather than synthetic drug-like molecules. The Spacial score (SPS)[10] sums per-atom contributions from hybridization, stereochemistry, ring fusion, and substitution, and therefore responds directly to three-dimensional character. Böttcher's score[11] is an information-theoretic measure that accumulates the microenvironmental diversity of each atom and consequently scales with molecular size as well as with structural intricacy.

CbyD targets are shifted toward higher complexity than USPTO-190[5] and PaRoutes[4] targets on SAScore, NPScore, and SPS, with the separation most pronounced for NPScore, consistent with the predominance of natural-product targets in CbyD. The Böttcher's score distributions overlap substantially across all three datasets, which is expected given that the metric scales with molecular size and that the molecular-weight distributions of the three datasets largely coincide.

The conclusion drawn in the main text therefore does not rest on any single metric, and the metric most sensitive to size is also the one that discriminates least between the datasets.

### S3.2 Reaction complexity

This section reports the analyses supporting the comparison of CbyD single-step reactions with USPTO-Full[6] and Pistachio[7] in the main text; the number bonds formed per reaction, the complexity of the reaction products, representative examples of the mixed reaction classes discussed there (**Fig. S8**), and the distribution of solvent usage within CbyD.

Counting the bonds formed in each reaction separates skeleton construction from functional-group manipulation (**Fig. S6**). The distribution of C–C bond formations in CbyD is shifted toward reactions forming a larger number of C–C bonds compared with USPTO-Full and Pistachio. This difference is consistent with the greater prevalence of bond-forming steps that build or connect larger molecular fragments in published total syntheses. The pattern is reversed for C–X bonds: C–X bond formation is present in a majority of USPTO-Full and Pistachio reactions but occurs in a smaller fraction of CbyD reactions. This is consistent with the reaction-class distributions presented in the main text, in which CbyD is enriched in C–C bond formation and depleted in the heteroatom alkylation, arylation, and acylation classes that dominate the two large-scale datasets.

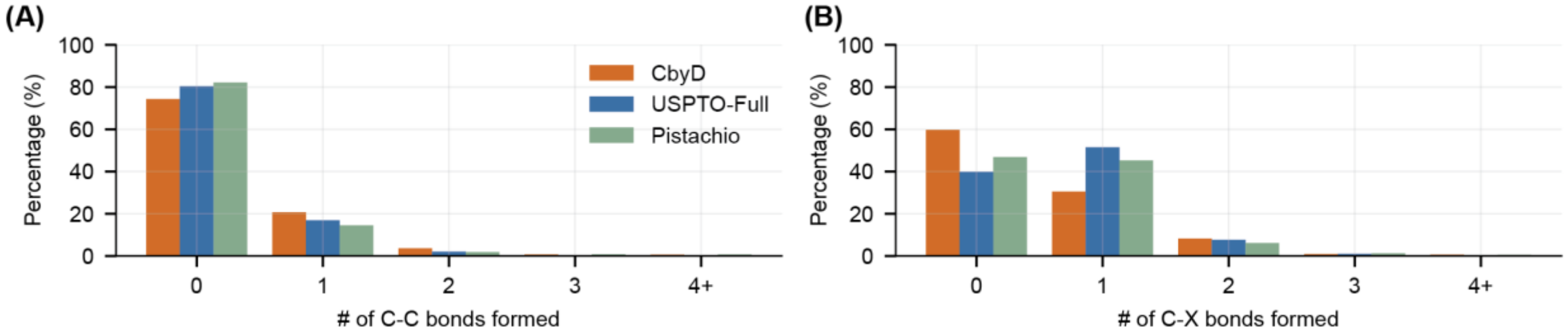

**Figure S6. (A)** Distribution of the number of C–C and **(B)** C–X bonds formed per reaction in CbyD, USPTO-Full, and Pistachio datasets.

Additionally, CbyD products of single reactions are generally more complex (**Fig. S7**). Most USPTO-Full and Pistachio products contain no tetrahedral stereocenters (**Fig. S7G**), while CbyD products are distributed over a broad range extending beyond ten. The intermediates encountered along a total synthesis are thus not appreciably larger than those in the conventional datasets, but are substantially richer in stereochemical and three-dimensional content, mirroring the difference observed for the final targets.

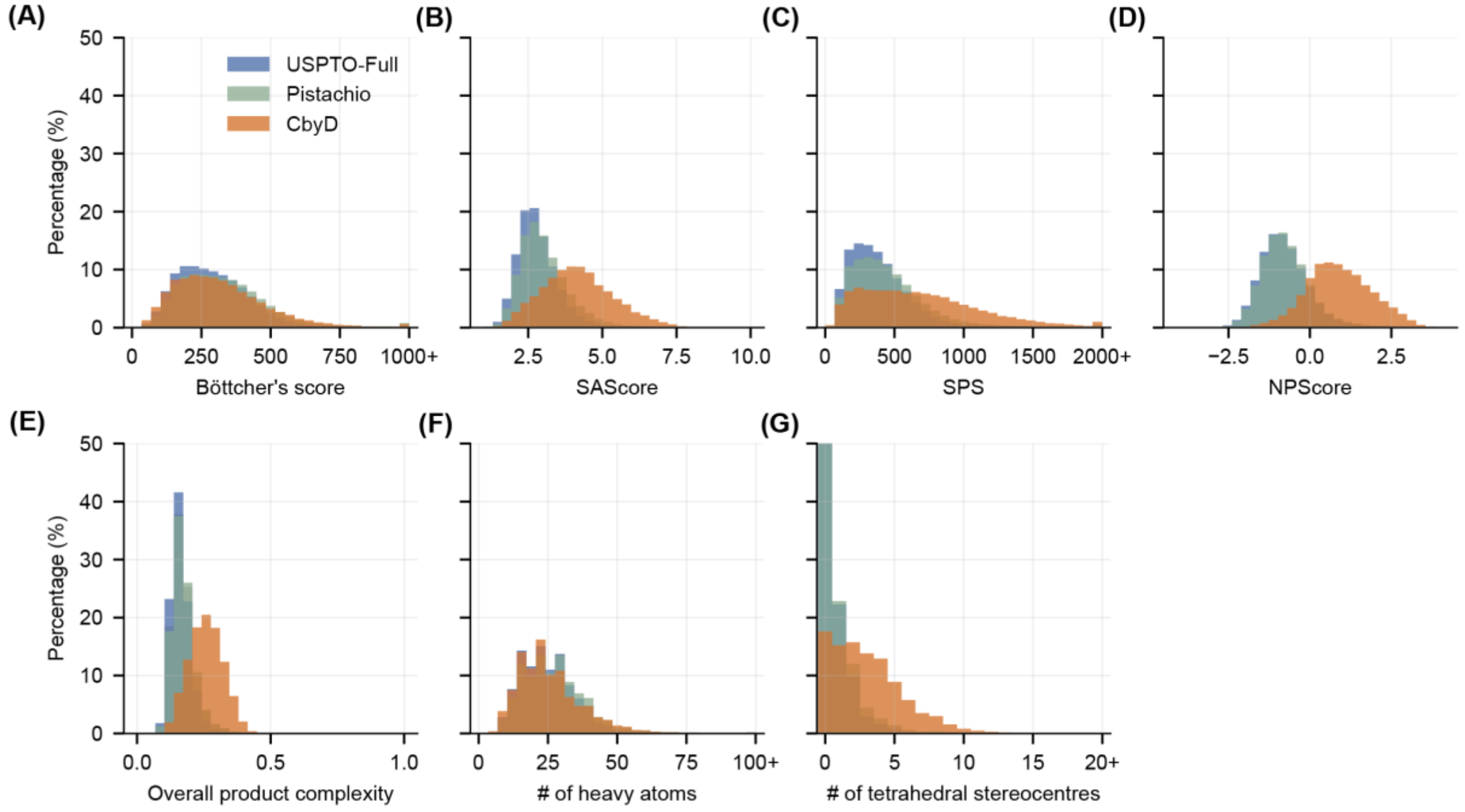

**Figure S7.** Distribution of molecular complexity metrics for reaction products in CbyD, USPTO-Full, and Pistachio. **(A)** Böttcher's score. **(B)** SAScore**.** **(C)** SPS. **(D)** NPScore. **(E)** Overall product complexity, defined as the average of the normalized metrics in **(A)–(D)**. **(F)** # of heavy atoms. **(G)** # of tetrahedral stereocenters.

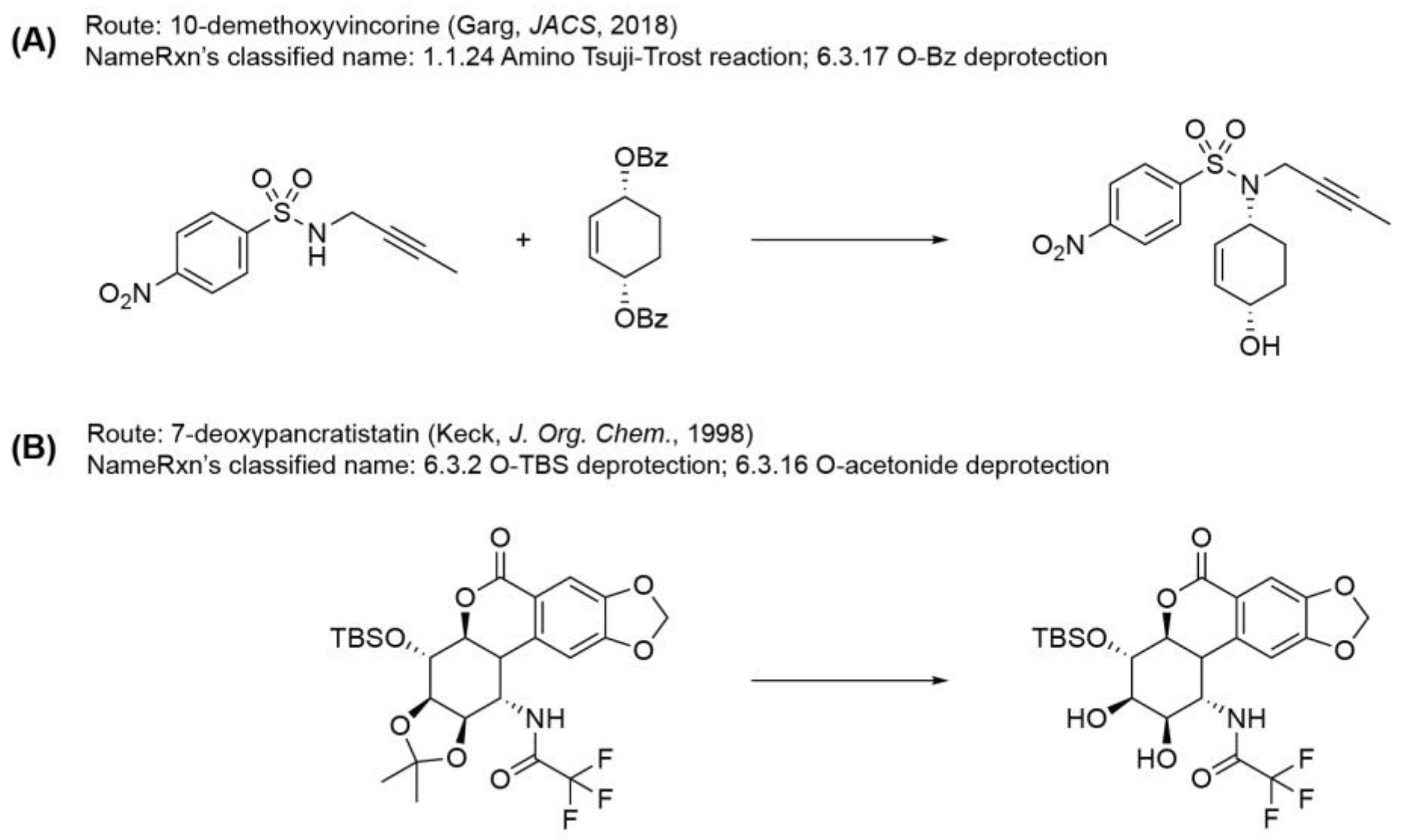

**Figure S8.** **(A)** Example of "mixed first-level classes" reaction. **(B)** Example of "mixed second-level classes" reaction.

**Fig. S9** shows the ten solvents most frequently used across CbyD reactions, ranked by the number of reactions in which each appears. This analysis is possible for CbyD alone, because it is the only dataset in the comparison for which solvents are explicitly annotated as such rather than pooled with the other reaction components.

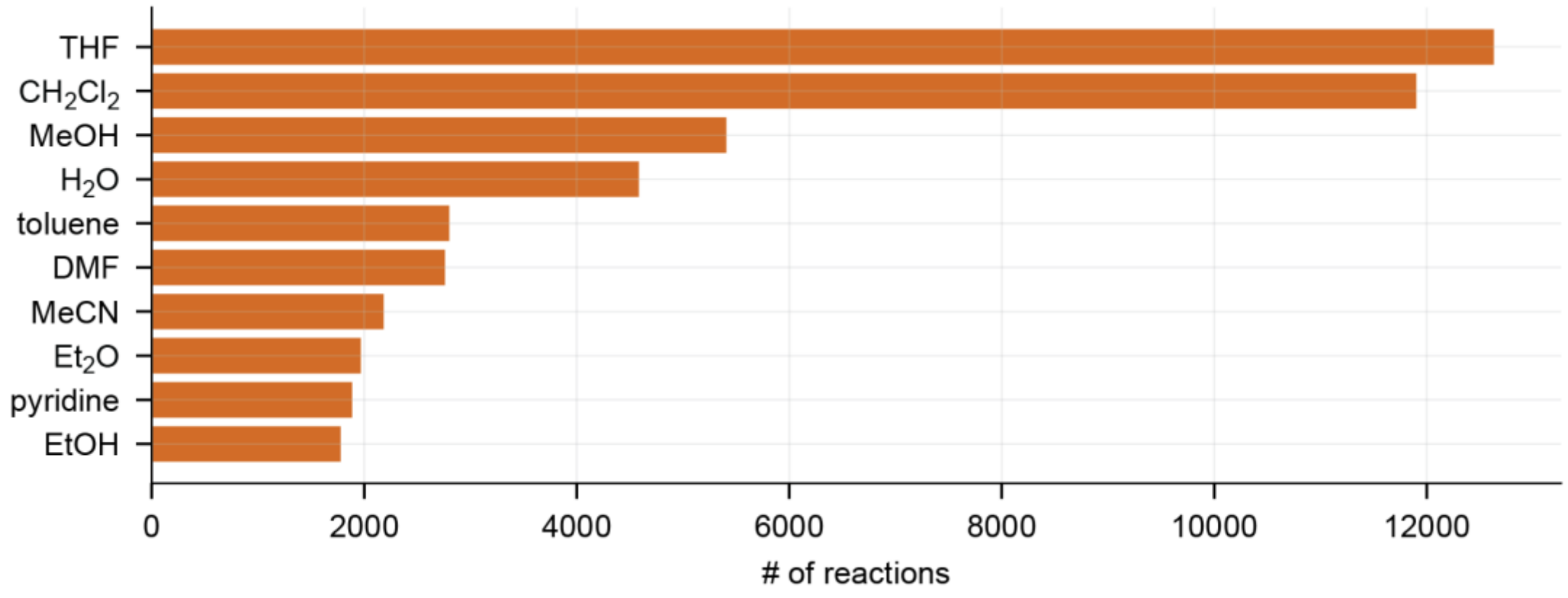

**Figure S9.** The ten most frequently used solvents in CbyD, ranked by number of reactions.

We enumerated the nitrogen- and oxygen-containing ring systems present in the final target structures of all CbyD routes; intermediates and building blocks were not included. Each unique ring system was counted once per target, and counts are reported as the number and percentage of targets containing that ring. Among nitrogen heterocycles (**Fig. S10A**), pyridine, piperidine, and indole, are the most frequent, followed by pyrrolidine, pyrrole, and piperazine. Among oxygen heterocycles (**Fig. S10B**), tetrahydropyran and tetrahydrofuran are the most frequent, followed by epoxide, butenolide, and γ-butyrolactone. For comparison, the heterocycle compositions of FDA-approved small-molecule drugs have been reported for N- and O-heterocycles.[18–20] The most frequent ring systems are largely shared (pyridine and piperidine, tetrahydropyran and tetrahydrofuran) but they occur far less often in CbyD targets.

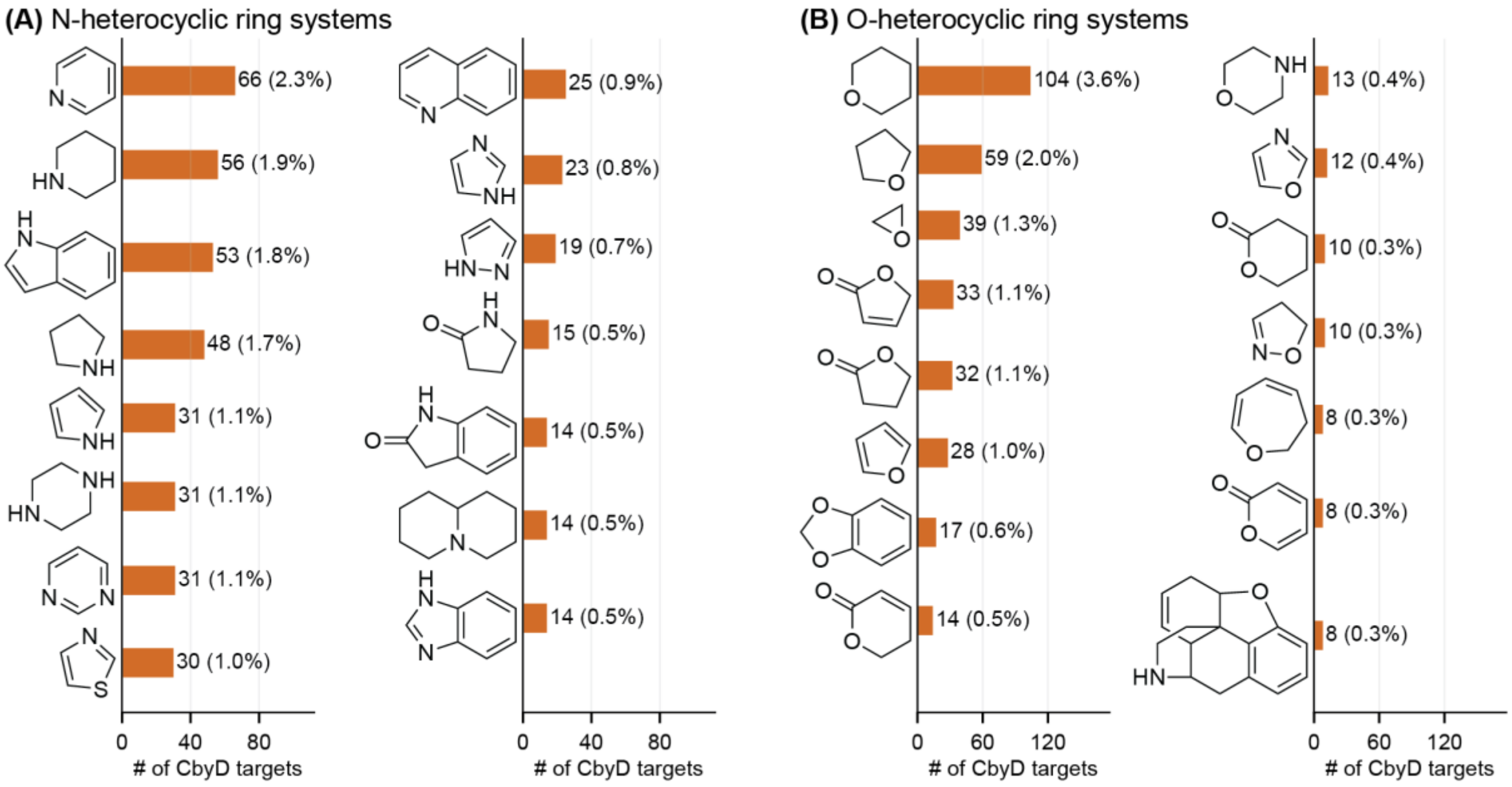

**Figure S10.** Most frequent **(A)** nitrogen and **(B)** oxygen heterocycles among the 2,902 unique CbyD target molecules (each ring system counted once per target), ranked by the number of targets containing each ring system.

### S3.3 Starting-material accessibility

We examined the starting materials used in CbyD routes, focusing on their structural characteristics, commercial availability, and role in establishing stereochemistry. We found the following:

1. We compared chiral starting materials in CbyD, USPTO-190,[5] and PaRoutes[4] and observed substantial differences in their structural complexity and apparent availability. Chiral starting materials in PaRoutes are predominantly small building blocks, such as protected amino acids, amino alcohols, and cyclic amines, that are consistent with commercially available synthetic reagents (**Fig. S11B**). In contrast, many starting materials assigned in USPTO-190 are structurally complex and densely functionalized molecules (**Fig. S11A**). These are unlikely to represent actual purchasable starting materials and may instead correspond to intermediates at which route reconstruction was truncated. The chiral starting materials in CbyD are generally smaller and more consistent with commonly used chiral building blocks in synthetic chemistry, including those frequently encountered across published synthesis routes.[21]

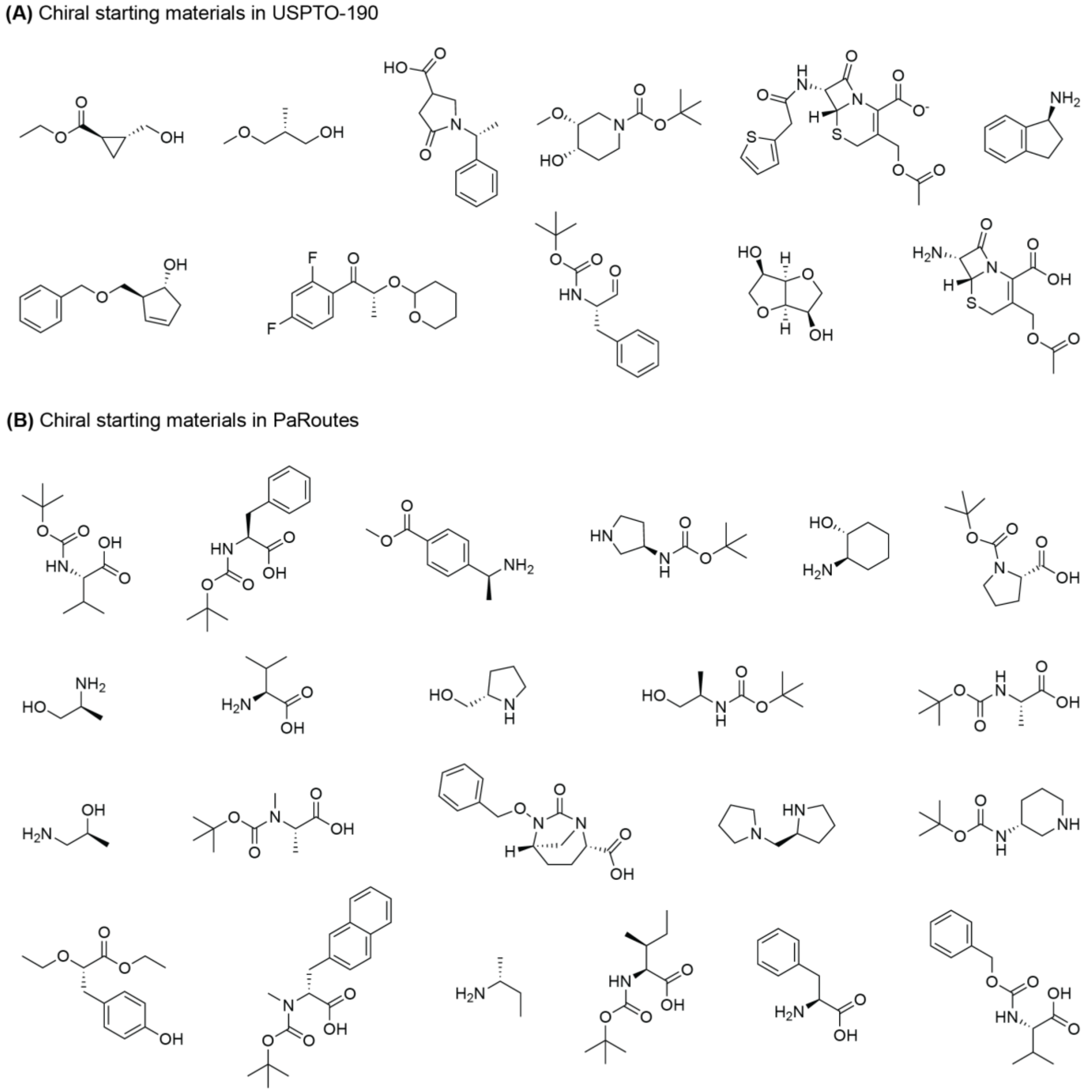

**Figure S11.** Examples of chiral starting materials in (A) USPTO-190 and (B) PaRoutes.

2. We also next assessed the accessibility of CbyD starting materials using an exemplary open-access dataset of building blocks from eMolecules (free download of 1 September 2026, 33.6 million SMILES).[22] Among all unique starting materials, 64.9% were identified in their exact stereochemical form in this catalogue. We then applied chemical normalization to account for differences in representation, covalent/ionic reagent representations, phosphorus ylide representations, and counterion-agnostic matching. After normalizing, the fraction identified as purchasable increases to 71.0%. The remaining 29.0% were not identified (**Fig. S12**). We further examined the reasons for these unmatched starting materials. For 2.6% of all starting materials, a different stereoisomer of the same compound was available, while 26.4% had no corresponding scaffold identified in the databases. The unmatched compounds were predominantly carbon-containing molecules (95.9%), with the remaining 4.1% consisting of inorganic or small non-carbon reagents such as $H_2O$, $H_2O_2$, $I_2$, $Br_2$, and $OsO_4$. It is notable that some common carbon-containing reagents, such as methyl iodide, are absent from this download file even though they can be found through the eMolecules website. Here, we use eMolecules as one example to illustrate that the choice of commercial compound database can affect the assessment of starting-material availability in computer-aided synthesis planning.

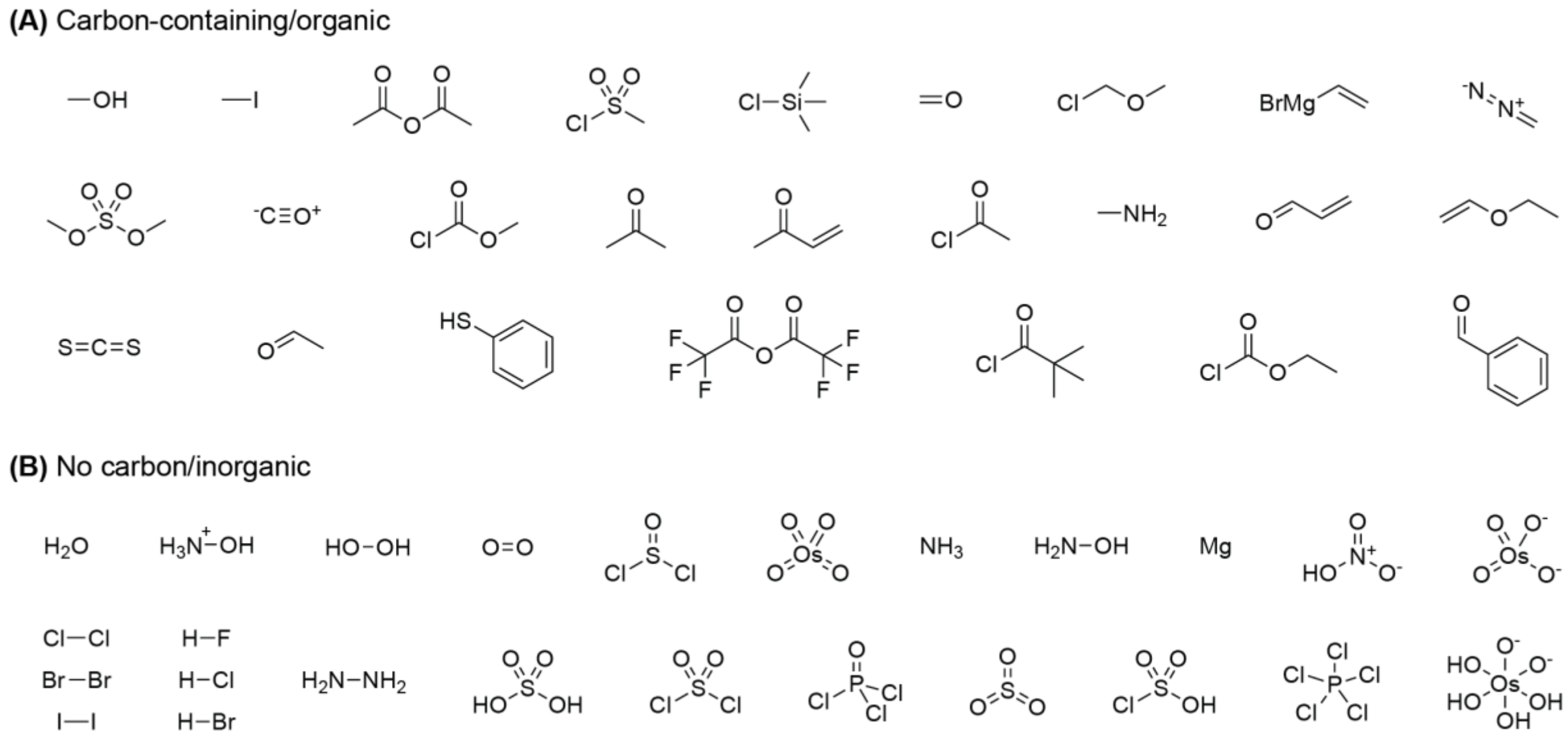

**Figure S12.** Examples of starting materials and reagents in CbyD that are *not* in the open-access eMolecules download file of building blocks.

3. Stereochemistry in a synthetic route can be established either by stereoselective reactions or by incorporating stereocenters from chiral starting materials. To examine how CbyD routes establish stereochemistry, we traced each target stereocenter backward through the route to the reaction that formed it or to the corresponding stereocenter in a starting material (**Fig. S13**). We found that 83% of target stereocenters are formed during the route, while 17% are directly inherited from chiral starting materials. Thus, most stereocenters in CbyD are formed through reactions, providing substantial representation of stereoselective transformations. It is also note that resolution and purification steps are not consistently represented as separate reactions in CbyD, making it difficult to reliably determine the fraction of racemic and enantioselective syntheses. This 17% is a lower bound on the contribution of chiral starting materials to stereochemical control. Some stereocenters classified as formed during the route are established using chiral auxiliaries or catalysts that originate from chiral starting materials. Because these reagents are subsequently removed, their stereochemical contribution cannot be traced to the target structure. Two representative examples are shown in **Fig. S13**. In the podophyllotoxin route, all four target stereocenters are classified as formed during the route, but their stereochemical information is derived from a valine-derived oxazolidinone that is subsequently cleaved. The preussin route illustrates a mixed case: one stereocenter is directly inherited from L-phenylalanine and subsequently directs the diastereoselective formation of two additional stereocenters.

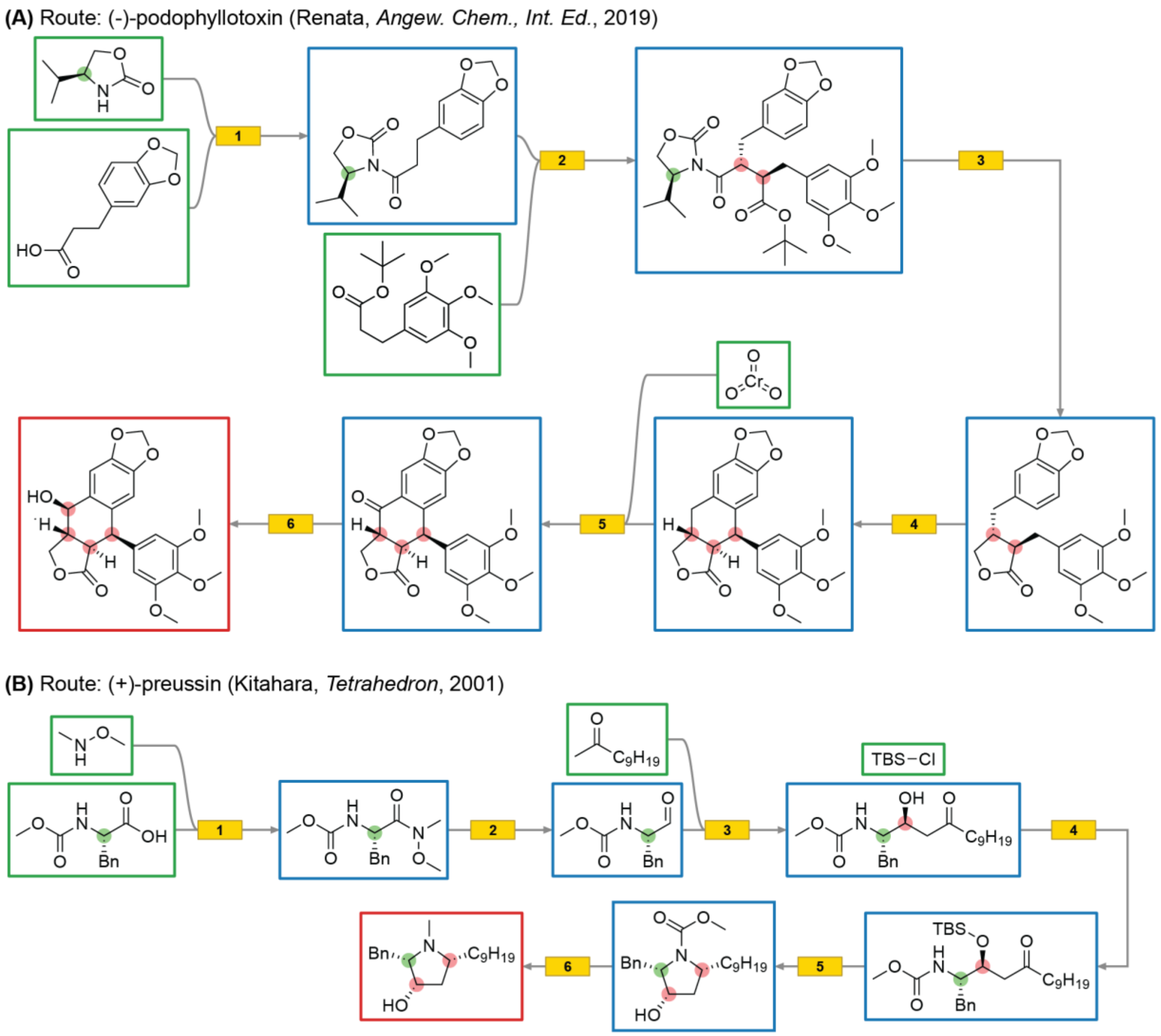

**Figure S13.** Each target stereocenter is traced back through the route, one reaction at a time, to either the step that creates it (installed, red) or a stereocenter already present in a purchased starting material (purchased, green). **(A)** All four target stereocenters are installed. The only chiral starting material, the valine-derived oxazolidinone auxiliary, sets the first two stereocenters through a diastereoselective coupling and is then cleaved, so its stereocenter does not reach the target. **(B)** The benzylic stereocenter carries through unchanged from starting material. The carbinol and C5 stereocenters are installed by substrate-controlled aldol addition and cyclization.

### S3.4 Key step analysis

There is no universally accepted definition for a key step in a total synthesis. Here, we identify key steps as reactions that produce the largest increase in molecular complexity. This is measured with several complexity metrics. Key steps are identified in the following steps.[23]

1. Reactions are classified by NameRxn.[16] Protection, deprotection and redox are discarded. Formation of cyclic acetals, ketals and dithianes is also discarded since these are protecting groups classified outside protection class.
2. The change in complexity is $\Delta C = C_{product} - \sum C_{reactant}$ for additive metrics (Böttcher's score, SPS). It is $\Delta C = C_{product} - max(C_{reactant})$ for per-molecule metrics (SAScore, NPScore), because these scores lie on a fixed scale independent of molecule size.
3. Among the reactions not excluded in step 1, a reaction is defined as a key step if its $\Delta C$ falls within the top 10% of scores across the dataset (**Table S14**).

We applied this procedure separately to each of the four complexity metrics, including Böttcher's score,[11] SAScore,[8] SPS,[10] NPScore.[9] For each metric, we tracked the complexity of intermediates along the longest linear sequence of each route, from starting material to target. We then classified the identified key steps by the NameRxn[16] superclass and counted the most frequent individual reactions (**Fig. S14**).

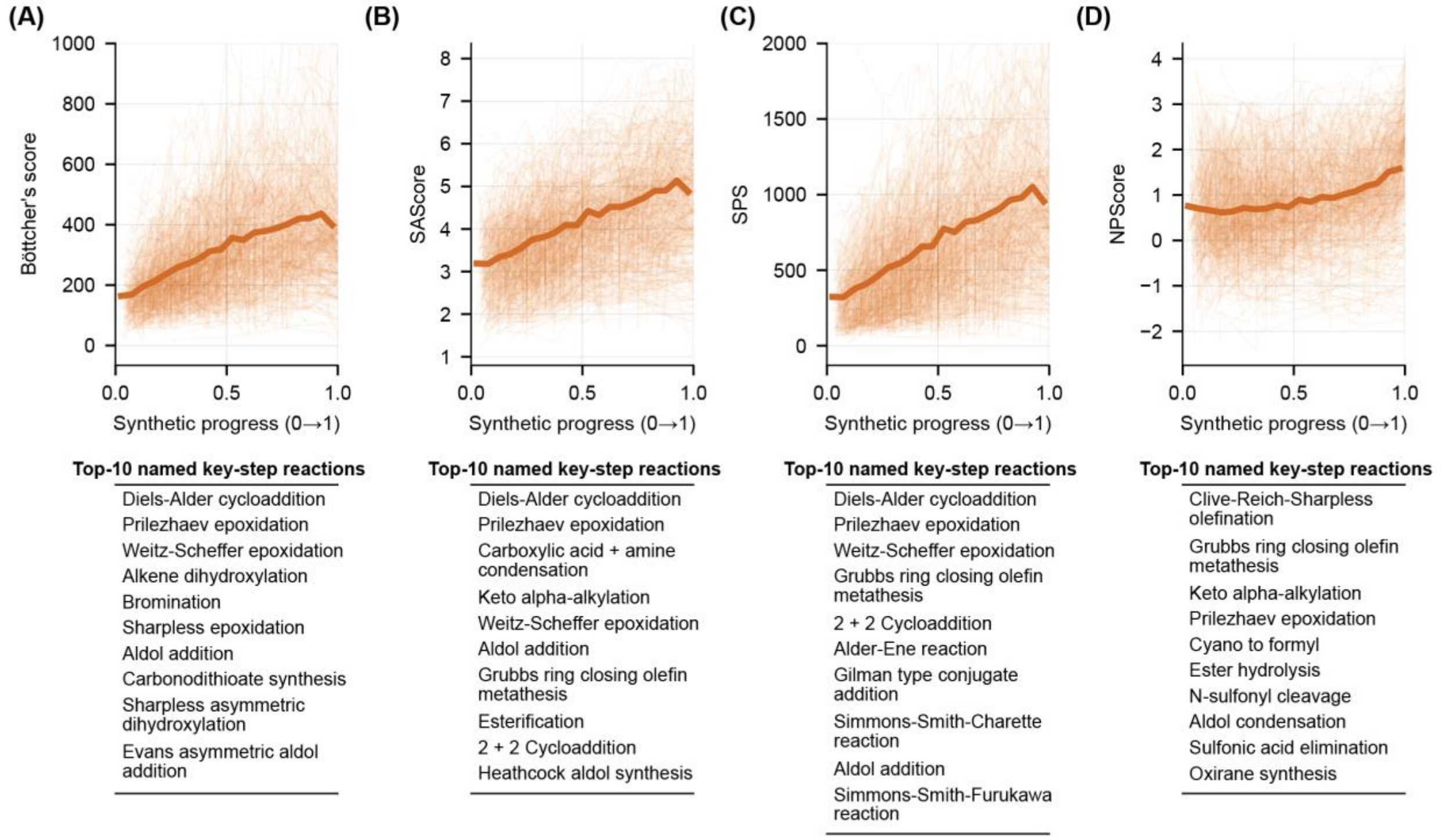

**Figure S14**. **(Top)** Molecular complexity along CbyD routes under four metrics: **(A)** Böttcher's score, **(B)** SAScore, **(C)** SPS, **(D)** NPScore, plotted against normalized synthetic progress. Thin lines show individual routes, and the thick line shows the mean trend. **(Bottom)** The 10 most frequent named reactions among key steps, identified under each metric.

The first three give similar complexity profiles, rising roughly two- to three-fold and monotonically with synthetic progress, and their top-ranked key steps overlap substantially: Diels–Alder cycloaddition is the most frequent in all three, followed by epoxidation. The differences between them reflect the unique strengths of each metric. Böttcher's score, which accounts for heteroatom diversity, favors oxidative and heteroatom-installing steps such as alkene dihydroxylation, bromination, and xanthate formation. SAScore, based on fragment rarity, favors polar two-component couplings such as esterification and carboxylic acid–amine condensation. SPS, which weights stereochemistry and $sp^3$ character, favors stereocenter-forming cyclizations such as the [2+2] cycloaddition, the Alder–ene reaction, and Simmons–Smith cyclopropanation. NPScore behaves differently: its profile is nearly flat (approximately 0.7 to 1.6) with a spread that includes negative values, and its top-ranked steps are dominated by functional-group interconversions and deprotections, such as cyano-to-formyl conversion, *N*-sulfonyl cleavage, ester hydrolysis, and sulfonic acid elimination, with the Diels–Alder cycloaddition absent entirely. This is expected, since NPScore quantifies resemblance to natural-product chemical space rather than structural complexity, and is therefore not a suitable basis for key-step identification. Because Böttcher's score, SAScore, and SPS identify largely overlapping key steps, we show SAScore in the main text (**Fig. 7**) and in the route insets (**Figs. S15–S19**).

**Figs. S15–S19** show the full route trees with highlighted key steps. The SAScore of each molecule along the longest linear sequence is plotted in the inset. In most routes, the early steps modify functional groups on simple aromatic or monocyclic precursors and change the SAScore by less than one unit per step. The largest increases often occur at steps that form several rings or stereocenters at once.

- (±)-isocomene, Wender 1981[24] (**Fig. S15**): at step 4, an intramolecular [3+2] *meta*-photocycloaddition forms the key tetracyclic intermediate.
- (+)-isomatrine, Reisman 2022[25] (**Fig. S16**): at step 1, a pivotal dearomative annulation of pyridine and glutaryl chloride rapidly constructs the tetracyclic core.
- (–)-longifolene, Sarpong 2022[26] (**Fig. S17**): at step 2, a radical, titanium-mediated 4-*exo-trig* cyclization of epoxycarvone forms the desired bicyclo[3.1.1]heptanol.
- penicibilaene A, Dong 2021[27] (**Fig. S18**): at step 5, a directed Rh-catalyzed C–C activation allows for a "cut-and-sew" reaction with a trisubstituted alkene to forge the tricyclic core and all-carbon quaternary center.
- kopsinine, Boger 2013[28] (**Fig. S19)**: at step 8, a thermal cycloaddition cascade consisting of a [4+2] aza-Diels-Alder followed by [3+2] dipolar cycloaddition is used to construct the pentacyclic skeleton and set five contiguous stereocenters.

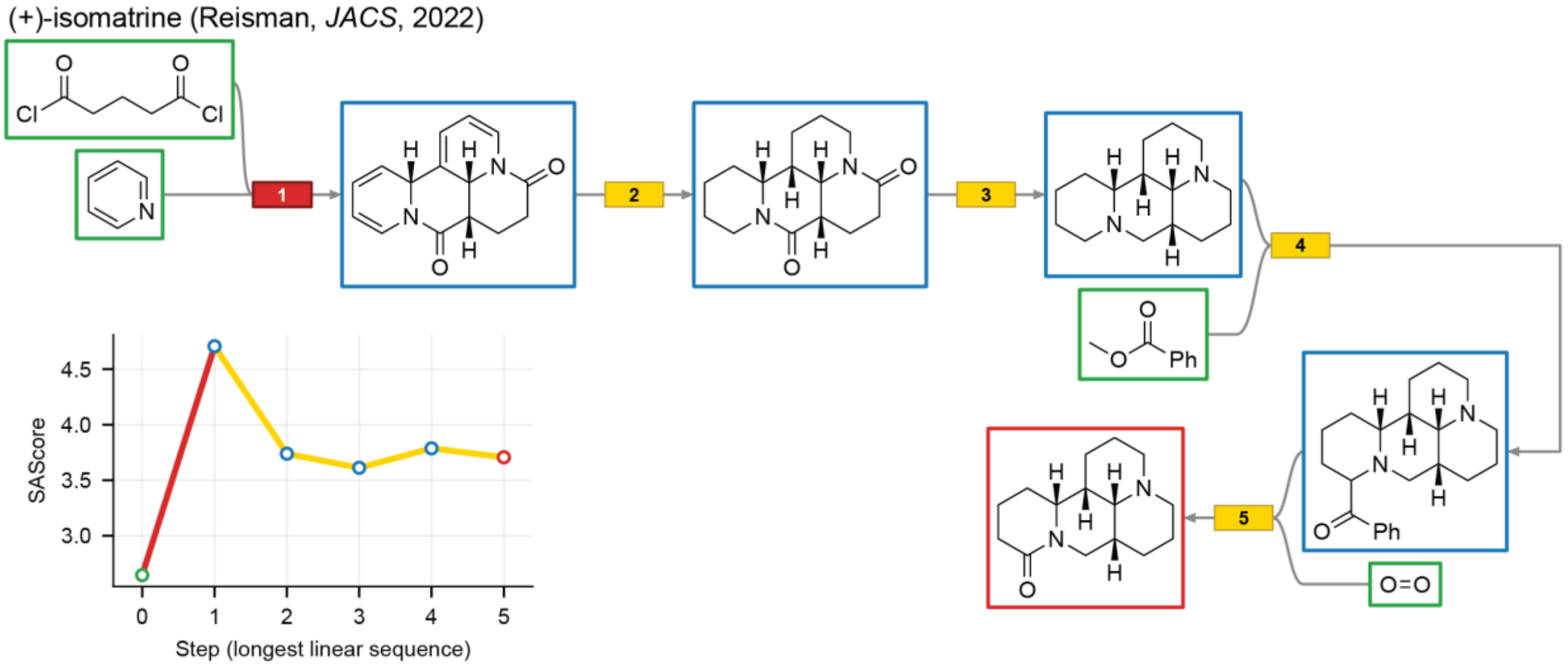

**Figure S15**. Example CbyD routes with annotated key step. Green boxes are starting materials, blue boxes are intermediates, and red boxes are targets. Yellow nodes are reactions, and red nodes are reactions annotated as key steps. Double edges indicate two equivalents of a reactant. Insets show the SAScore of each molecule along the longest linear sequence. Points are colored as their boxes, and red segments mark a key complexity-building step.

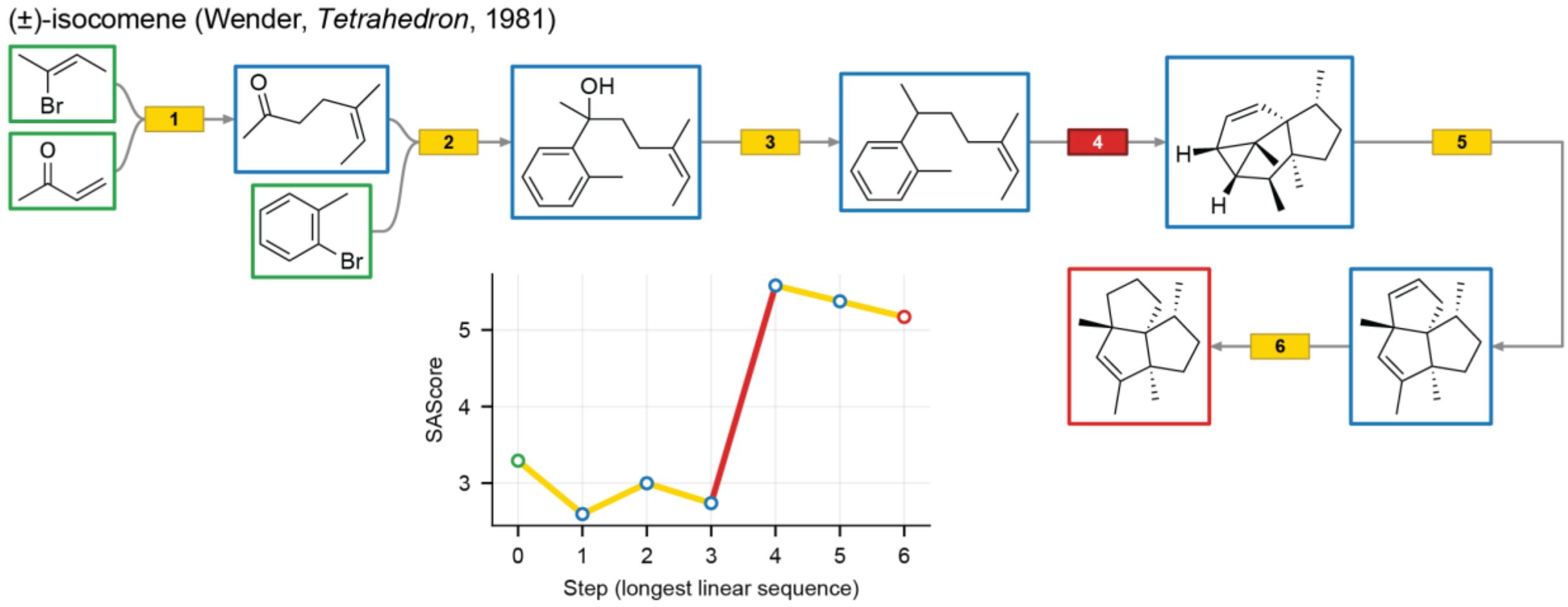

**Figure S16**. Example CbyD routes with annotated key step. Same format as **Figure S15**.

(-)-longifolene (Sarpong, *Nat. Chem.* 2022)

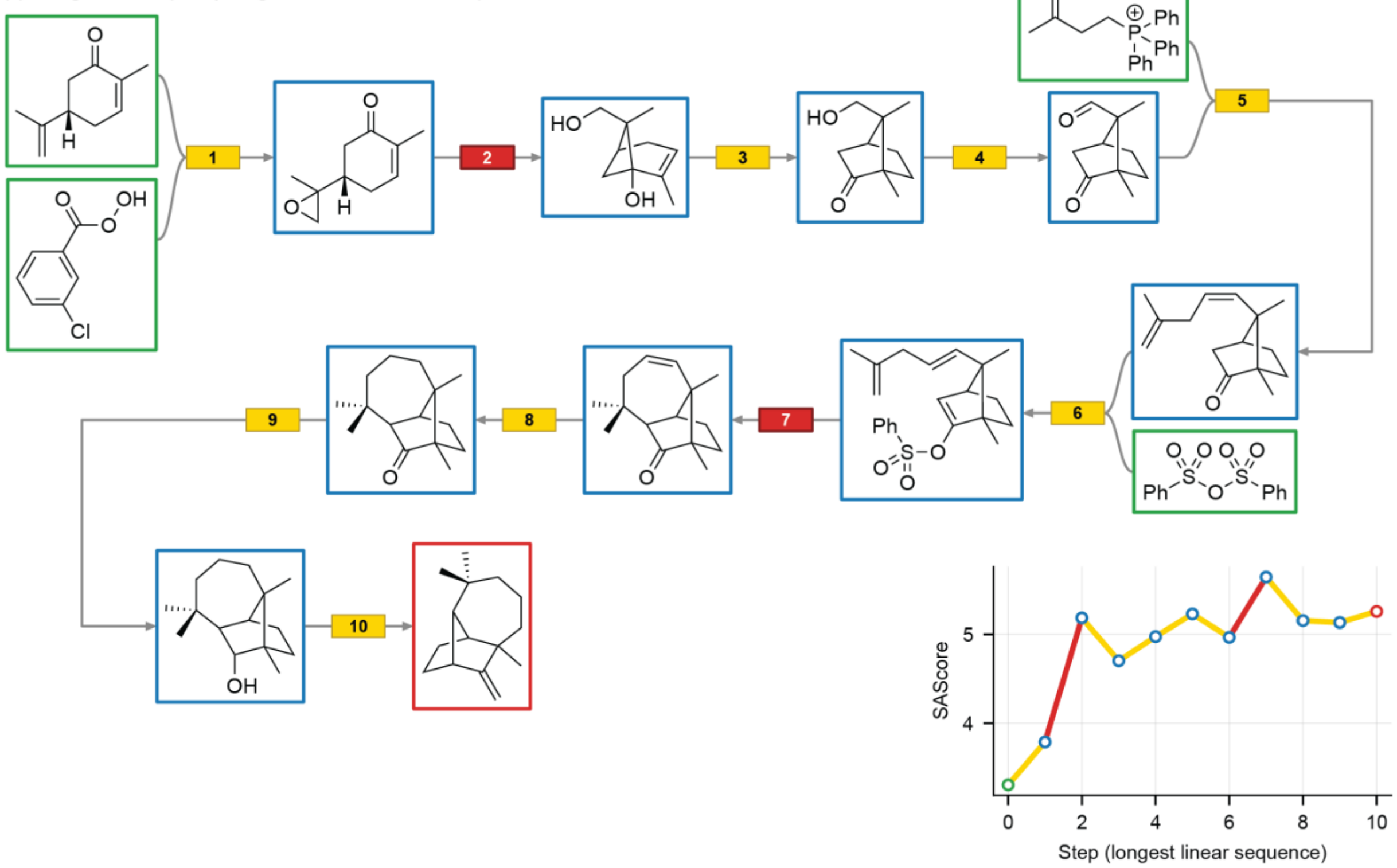

**Figure S17**. Example CbyD routes with annotated key step. Same format as **Figure S15**.

penicibilaene A (Dong, *JACS*, 2021)

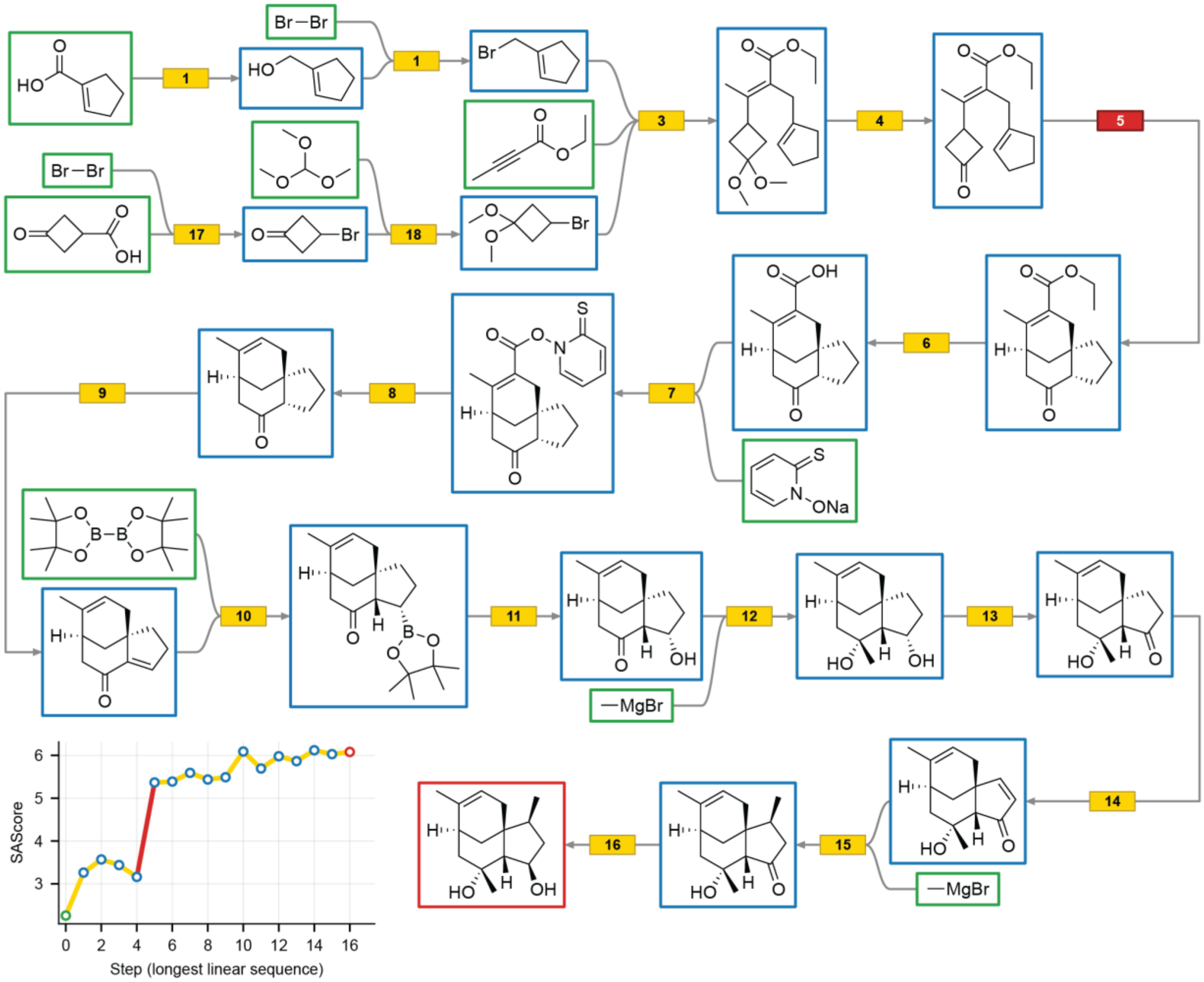

**Figure S18.** Example CbyD routes with annotated key step. Same format as **Figure S15**.

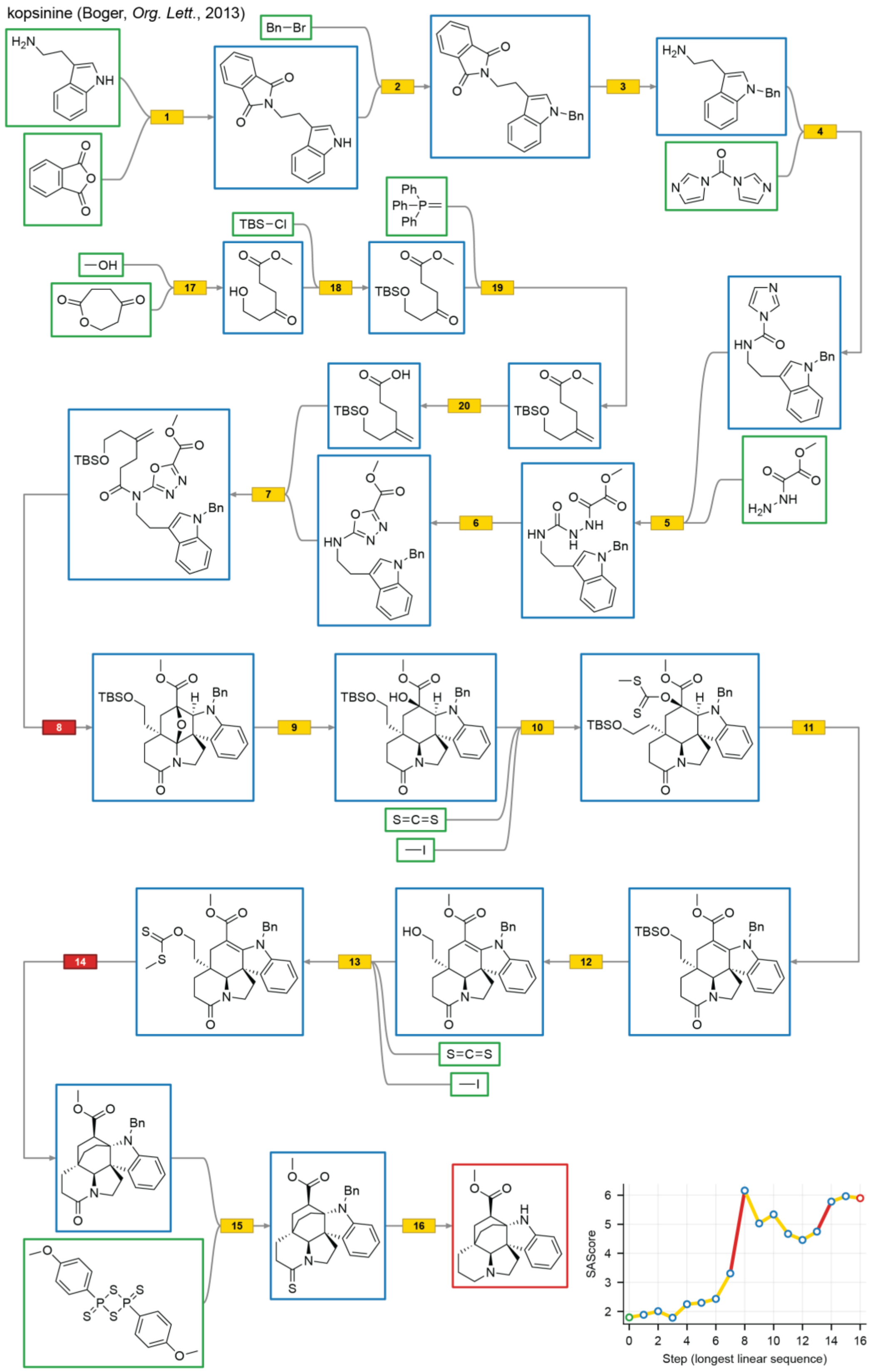

**Figure S19.** Example CbyD routes with annotated key step. Same format as **Figure S15**.

### S3.5 Supplementary tables

**Table S1**. Number of CbyD routes by publication year. Routes published in or before 1940 are pooled in the first row, as in Fig. 3A; the table covers all 3,502 routes.

| Year | Routes | Year | Routes | Year | Routes | Year | Routes |
|---|---|---|---|---|---|---|---|
| ≤1940 | 44 | 1962 | 2 | 1984 | 16 | 2006 | 57 |
| 1941 | 1 | 1963 | 7 | 1985 | 24 | 2007 | 78 |
| 1942 | 2 | 1964 | 3 | 1986 | 15 | 2008 | 75 |
| 1943 | 1 | 1965 | 5 | 1987 | 22 | 2009 | 74 |
| 1944 | 2 | 1966 | 3 | 1988 | 17 | 2010 | 117 |
| 1945 | 1 | 1967 | 8 | 1989 | 26 | 2011 | 136 |
| 1946 | 1 | 1968 | 8 | 1990 | 23 | 2012 | 154 |
| 1947 | 1 | 1969 | 6 | 1991 | 16 | 2013 | 134 |
| 1948 | 4 | 1970 | 14 | 1992 | 16 | 2014 | 117 |
| 1949 | 1 | 1971 | 12 | 1993 | 23 | 2015 | 106 |
| 1950 | 1 | 1972 | 14 | 1994 | 24 | 2016 | 148 |
| 1951 | 2 | 1973 | 12 | 1995 | 15 | 2017 | 129 |
| 1952 | 2 | 1974 | 12 | 1996 | 36 | 2018 | 141 |
| 1953 | 2 | 1975 | 15 | 1997 | 26 | 2019 | 144 |
| 1954 | 2 | 1976 | 16 | 1998 | 22 | 2020 | 141 |
| 1955 | 2 | 1977 | 11 | 1999 | 46 | 2021 | 164 |
| 1956 | 2 | 1978 | 15 | 2000 | 49 | 2022 | 208 |
| 1957 | 3 | 1979 | 20 | 2001 | 43 | 2023 | 107 |
| 1958 | 3 | 1980 | 18 | 2002 | 65 | 2024 | 135 |
| 1959 | 2 | 1981 | 20 | 2003 | 41 | 2025 | 37 |
| 1960 | 5 | 1982 | 23 | 2004 | 78 | 2026 | 47 |
| 1961 | 1 | 1983 | 14 | 2005 | 67 | | |

**Table S2.** Number of CbyD targets synthesized by a given number of routes (**Fig. 3B**). Targets are grouped by name after removal of stereodescriptor prefixes such as (+)-, (−)- and (±)-, giving 2,591 target names in this analysis. The 2,902 unique targets reported elsewhere are counted by unique SMILES.

| **Routes per target** | **1** | **2** | **3** | **4** | **5** | **6** | **7** | **8** | **9** | **10** | **11** | **12** | **17** | **23** | **25** | **31** |
|---|---|---|---|---|---|---|---|---|---|---|---|---|---|---|---|---|
| **Targets** | 2,196 | 237 | 71 | 30 | 13 | 12 | 8 | 4 | 6 | 2 | 3 | 4 | 1 | 2 | 1 | 1 |

**Table S3.** Summary statistics of target properties of CbyD, USPTO-190 and PaRoutes (**Fig. 3C**), one entry per unique target.

| Dataset | *n* | Mean | SD | Median | Q1 | Q3 | Min | Max |
|---|---|---|---|---|---|---|---|---|
| **Overall target complexity** | | | | | | | | |
| CbyD | 2,902 | 0.36 | 0.10 | 0.37 | 0.29 | 0.43 | 0.09 | 0.74 |
| USPTO-190 | 190 | 0.24 | 0.05 | 0.24 | 0.20 | 0.27 | 0.13 | 0.54 |
| PaRoutes | 10,000 | 0.17 | 0.05 | 0.17 | 0.14 | 0.20 | 0.06 | 0.80 |
| **Molecular weight** | | | | | | | | |
| CbyD | 2,902 | 385.67 | 155.87 | 354.49 | 288.39 | 448.58 | 60.05 | 1,875.31 |
| USPTO-190 | 190 | 458.63 | 170.22 | 417.86 | 345.85 | 518.79 | 181.67 | 956.02 |
| PaRoutes | 10,000 | 378.24 | 135.07 | 365.41 | 282.17 | 454.52 | 123.11 | 2,849.32 |
| **Tetrahedral stereocenters** | | | | | | | | |
| CbyD | 2,902 | 4.24 | 3.14 | 4.00 | 2.00 | 6.00 | 0.00 | 30.00 |
| USPTO-190 | 190 | 1.97 | 1.69 | 2.00 | 1.00 | 3.00 | 0.00 | 14.00 |
| PaRoutes | 10,000 | 0.53 | 1.06 | 0.00 | 0.00 | 1.00 | 0.00 | 17.00 |
| **SAScore** | | | | | | | | |
| CbyD | 2,902 | 4.71 | 1.28 | 4.68 | 3.78 | 5.62 | 1.14 | 8.31 |
| USPTO-190 | 190 | 3.57 | 0.64 | 3.55 | 3.12 | 3.96 | 2.39 | 6.96 |
| PaRoutes | 10,000 | 2.74 | 0.61 | 2.65 | 2.33 | 3.05 | 1.37 | 8.99 |
| **NPScore** | | | | | | | | |
| CbyD | 2,902 | 1.46 | 1.23 | 1.66 | 0.80 | 2.32 | -2.23 | 4.01 |
| USPTO-190 | 190 | -0.41 | 0.72 | -0.53 | -0.89 | -0.02 | -2.01 | 2.46 |
| PaRoutes | 10,000 | -0.87 | 0.67 | -0.93 | -1.34 | -0.46 | -2.92 | 2.84 |
| **SPS** | | | | | | | | |
| CbyD | 2,902 | 891.65 | 445.75 | 828.00 | 564.00 | 1,142.00 | 24.00 | 3,890.00 |
| USPTO-190 | 190 | 651.57 | 270.29 | 620.00 | 499.25 | 767.50 | 147.00 | 2,253.00 |
| PaRoutes | 10,000 | 390.46 | 206.60 | 356.00 | 245.00 | 492.00 | 74.00 | 3,097.00 |
| **Böttcher score** | | | | | | | | |
| CbyD | 2,902 | 384.21 | 177.49 | 360.30 | 274.23 | 455.84 | 10.75 | 2,046.74 |
| USPTO-190 | 190 | 387.80 | 138.88 | 384.40 | 280.18 | 458.03 | 93.36 | 781.33 |
| PaRoutes | 10,000 | 296.21 | 128.36 | 280.06 | 204.64 | 370.17 | 45.17 | 3,018.13 |

**Table S4**. Distributions of overall target complexity, molecular weight and number of tetrahedral stereocenters as plotted in **Fig. 3C** (% of each dataset's targets per bin).

| Bin | CbyD | USPTO-190 | PaRoutes | Bin | CbyD | USPTO-190 | PaRoutes |
|---|---|---|---|---|---|---|---|
| **Overall target complexity** | | | | **Molecular weight (cont.)** | | | |
| 0–0.05 | 0.0 | 0.0 | 0.0 | 400–450 | 12.1 | 12.6 | 13.6 |
| 0.05–0.1 | 0.0 | 0.0 | 2.1 | 450–500 | 8.2 | 12.1 | 11.2 |
| 0.1–0.15 | 1.9 | 0.5 | 32.8 | 500–550 | 5.0 | 9.5 | 6.4 |
| 0.15–0.2 | 6.8 | 25.3 | 42.8 | 550–600 | 3.5 | 3.2 | 3.6 |
| 0.2–0.25 | 8.5 | 29.5 | 16.7 | 600–650 | 2.3 | 2.1 | 2.0 |
| 0.25–0.3 | 9.8 | 34.7 | 3.8 | 650–700 | 1.4 | 1.6 | 1.1 |
| 0.3–0.35 | 16.9 | 7.9 | 1.2 | 700–750 | 1.3 | 2.6 | 0.7 |
| 0.35–0.4 | 19.7 | 1.6 | 0.4 | 750–800 | 1.0 | 2.1 | 0.3 |
| 0.4–0.45 | 17.8 | 0.0 | 0.1 | 800–850 | 0.5 | 2.6 | 0.3 |
| 0.45–0.5 | 11.5 | 0.0 | 0.0 | 850–900 | 0.3 | 4.2 | 0.3 |
| 0.5–0.55 | 4.3 | 0.5 | 0.1 | 900–950 | 0.3 | 0.5 | 0.1 |
| 0.55–0.6 | 2.0 | 0.0 | 0.0 | 950–1000+ | 0.8 | 0.5 | 0.3 |
| 0.6–0.65 | 0.5 | 0.0 | 0.0 | **Tetrahedral stereocenters** | | | |
| 0.65–0.7 | 0.1 | 0.0 | 0.0 | 0 | 10.2 | 22.1 | 67.7 |
| 0.7–0.75 | 0.1 | 0.0 | 0.0 | 1 | 9.5 | 14.7 | 20.7 |
| 0.75–0.8 | 0.0 | 0.0 | 0.0 | 2 | 11.4 | 31.6 | 7.3 |
| 0.8–0.85 | 0.0 | 0.0 | 0.0 | 3 | 11.8 | 18.4 | 2.1 |
| 0.85–0.9 | 0.0 | 0.0 | 0.0 | 4 | 15.1 | 7.4 | 1.1 |
| 0.9–0.95 | 0.0 | 0.0 | 0.0 | 5 | 13.5 | 4.7 | 0.5 |
| 0.95–1 | 0.0 | 0.0 | 0.0 | 6 | 9.1 | 0.0 | 0.2 |
| **Molecular weight** | | | | 7 | 6.4 | 0.5 | 0.1 |
| 0–50 | 0.0 | 0.0 | 0.0 | 8 | 5.6 | 0.0 | 0.0 |
| 50–100 | 0.1 | 0.0 | 0.0 | 9 | 2.4 | 0.0 | 0.1 |
| 100–150 | 0.8 | 0.0 | 0.7 | 10 | 1.7 | 0.0 | 0.0 |
| 150–200 | 3.3 | 1.6 | 5.1 | 11 | 1.1 | 0.0 | 0.0 |
| 200–250 | 11.0 | 3.2 | 10.2 | 12 | 0.9 | 0.0 | 0.0 |
| 250–300 | 14.1 | 8.9 | 14.4 | 13 | 0.2 | 0.0 | 0.0 |
| 300–350 | 18.7 | 12.1 | 15.4 | 14 | 0.2 | 0.5 | 0.0 |
| 350–400 | 15.1 | 20.5 | 14.5 | 15+ | 1.0 | 0.0 | 0.0 |

**Table S5**. Ring junctions (**Fig. 3E**): % of targets with at least one fused (two rings sharing one bond), bridged (two rings sharing two or more bonds) or spiro (one atom shared by two ring systems) junction. A target can belong to several categories.

| **Ring junction** | **CbyD** | **USPTO-190** | **PaRoutes** |
|---|---|---|---|
| Fused | 76.7 | 54.2 | 43.2 |
| Bridged | 31.8 | 3.2 | 1.5 |
| Spiro | 6.7 | 5.3 | 1.6 |

**Table S6**. Ring types per target (**Fig. 3D**): % of each dataset's targets with the given number of rings of each type.

| Rings per target | CbyD | USPTO-190 | PaRoutes |
|---|---|---|---|
| **Aromatic rings** | | | |
| 1 | 25.7 | 20.5 | 23.0 |
| 2 | 17.7 | 29.5 | 30.3 |
| 3 | 9.0 | 31.1 | 24.3 |
| 4 | 3.5 | 8.9 | 12.4 |
| 5 | 1.4 | 3.7 | 3.0 |
| 6+ | 1.1 | 0.0 | 1.2 |
| **Heteroaromatic rings** | | | |
| 1 | 19.6 | 26.3 | 36.3 |
| 2 | 5.8 | 22.6 | 17.1 |
| 3 | 1.8 | 5.8 | 4.6 |
| 4 | 0.4 | 0.0 | 0.8 |
| 5 | 0.0 | 0.0 | 0.0 |
| 6+ | 0.1 | 0.0 | 0.0 |
| **Non-aromatic heterocycles** | | | |
| 1 | 26.7 | 54.2 | 30.1 |
| 2 | 22.5 | 16.3 | 7.2 |
| 3 | 13.5 | 2.1 | 0.9 |
| 4 | 5.0 | 0.0 | 0.0 |
| 5 | 1.9 | 0.0 | 0.0 |
| 6+ | 1.4 | 0.0 | 0.0 |
| **Non-aromatic carbocycles** | | | |
| 1 | 23.3 | 30.5 | 15.1 |
| 2 | 15.5 | 12.1 | 1.6 |
| 3 | 11.7 | 0.5 | 0.3 |
| 4 | 7.0 | 1.1 | 0.4 |
| 5 | 1.3 | 0.5 | 0.0 |
| 6+ | 0.9 | 0.0 | 0.0 |

**Table S7**. Percentage of targets containing at least one atom of each element or ring of each type, and number of targets per dataset (**Fig. 3D–F**).

| **Targets with** | **CbyD** | **USPTO-190** | **PaRoutes** |
|---|---|---|---|
| ≥1 N | 58.1 | 91.6 | 88.4 |
| ≥1 O | 93.0 | 94.7 | 90.5 |
| ≥1 S | 7.6 | 20.5 | 26.0 |
| ≥1 P | 0.7 | 0.5 | 0.7 |
| ≥1 Halogens | 13.3 | 56.8 | 48.0 |
| ≥1 aromatic rings | 58.4 | 93.7 | 94.2 |
| ≥1 heteroaromatic rings | 27.8 | 54.7 | 58.8 |
| ≥1 non-aromatic heterocycles | 71.0 | 72.6 | 38.3 |
| ≥1 non-aromatic carbocycles | 59.7 | 44.7 | 17.5 |
| acyclic | 1.3 | 0.0 | 2.3 |
| *n* targets | 2,902 | 190 | 10,000 |

**Table S8**. Heteroatoms per target (**Fig. 3F**): % of targets with the given number of atoms of each element.

| **Atoms per target** | **CbyD** | **USPTO-190** | **PaRoutes** | **Atoms per target** | **CbyD** | **USPTO-190** | **PaRoutes** |
|---|---|---|---|---|---|---|---|
| **O** | | | | **Halogens (F, Cl, Br, I)** | | | |
| 1 | 10.3 | 6.8 | 18.5 | 1 | 5.2 | 22.6 | 24.1 |
| 2 | 16.0 | 13.7 | 23.7 | 2 | 4.2 | 19.5 | 9.7 |
| 3 | 14.3 | 27.9 | 20.4 | 3 | 1.7 | 12.1 | 8.1 |
| 4 | 15.3 | 16.8 | 14.1 | 4 | 1.2 | 1.6 | 3.2 |
| 5 | 10.5 | 9.5 | 7.2 | 5 | 0.3 | 1.1 | 1.4 |
| 6 | 9.3 | 7.4 | 3.3 | 6+ | 0.6 | 0.0 | 1.6 |
| 7 | 5.7 | 4.2 | 1.7 | **S** | | | |
| 8 | 4.5 | 3.2 | 0.8 | 1 | 5.8 | 16.8 | 22.6 |
| 9 | 2.2 | 4.2 | 0.3 | 2 | 1.2 | 2.6 | 3.0 |
| 10+ | 4.8 | 1.1 | 0.5 | 3 | 0.2 | 0.0 | 0.4 |
| **N** | | | | 4 | 0.2 | 1.1 | 0.1 |
| 1 | 21.0 | 18.4 | 19.1 | 5 | 0.1 | 0.0 | 0.0 |
| 2 | 17.7 | 24.7 | 21.3 | 6+ | 0.1 | 0.0 | 0.0 |
| 3 | 6.5 | 16.8 | 17.7 | **P** | | | |
| 4 | 4.7 | 8.9 | 12.8 | 1 | 0.6 | 0.5 | 0.7 |
| 5 | 2.9 | 13.2 | 8.6 | 2 | 0.1 | 0.0 | 0.0 |
| 6 | 2.4 | 6.3 | 4.7 | 3 | 0.0 | 0.0 | 0.0 |
| 7 | 1.3 | 3.2 | 2.5 | 4 | 0.0 | 0.0 | 0.0 |
| 8 | 0.8 | 0.0 | 1.0 | 5 | 0.0 | 0.0 | 0.0 |
| 9 | 0.3 | 0.0 | 0.4 | 6+ | 0.0 | 0.0 | 0.0 |
| 10+ | 0.4 | 0.0 | 0.2 | | | | |

**Table S9**. NameRxn superclass distribution (**Fig. 4A**), as % of all reactions of each dataset. Reactions with several NameRxn assignments are mixed sub-classes (one superclass) or mixed classes (different superclasses). Class 11 (chiral and other separations, isomerizations, racemizations, purifications and salt formations) is removed from USPTO-Full and Pistachio during preprocessing from previous work.[14]

| Class | NameRxn superclass | CbyD | USPTO-Full | Pistachio |
|---|---|---|---|---|
| 0 | Unrecognized | 22.9 | 18.3 | 13.9 |
| 1 | Heteroatom alkylation/arylation | 7.5 | 21.9 | 19.7 |
| 2 | Acylation | 7.3 | 17.4 | 15.8 |
| 3 | C-C bond formation | 13.1 | 9.3 | 9.1 |
| 4 | Heterocycle formation | 3.3 | 3.8 | 3.1 |
| 5 | Protection | 4.8 | 0.7 | 1.7 |
| 6 | Deprotection | 9.7 | 10.9 | 13.6 |
| 7 | Reduction | 8.1 | 5.4 | 4.6 |
| 8 | Oxidation | 7.6 | 2.1 | 2.4 |
| 9 | Functional group interconversion | 9.1 | 7.6 | 9.2 |
| 10 | Functional group addition | 3.2 | 2.4 | 2.8 |
| 11 | Resolution and separation | 1.0 | 0.0 | 0.0 |
| mix sub-classes | Mixed second-level classes | 0.7 | 0.0 | 0.8 |
| mix classes | Mixed first-level classes | 1.7 | 0.0 | 3.4 |

**Table S10**. Reaction descriptors (**Fig. 4A** and **Fig. S6**): % of reactions with each number of reacting atoms, changing rings, stereochemical changes and C–C or C–X bonds formed.

| Value | CbyD | USPTO-Full | Pistachio |
|---|---|---|---|
| **Reacting atoms** | | | |
| 0 | 0.9 | 0.0 | 0.1 |
| 1 | 10.8 | 12.4 | 22.0 |
| 2 | 48.3 | 49.1 | 53.7 |
| 3 | 16.0 | 17.8 | 8.8 |
| 4 | 11.3 | 6.4 | 3.4 |
| 5 | 4.3 | 6.0 | 3.9 |
| 6+ | 8.5 | 8.2 | 8.0 |
| **Changing rings** | | | |
| 0 | 79.1 | 86.6 | 88.5 |
| 1 | 15.6 | 11.8 | 9.5 |
| 2 | 3.2 | 1.2 | 1.3 |
| 3 | 1.2 | 0.2 | 0.4 |
| 4+ | 0.9 | 0.1 | 0.2 |
| **Stereochemical changes (*R*/*S* + *E*/*Z*)** | | | |
| 0 | 67.9 | 95.2 | 93.8 |
| 1 | 19.4 | 3.0 | 4.1 |
| 2 | 7.4 | 1.4 | 1.5 |
| 3 | 3.0 | 0.2 | 0.3 |
| 4+ | 2.3 | 0.2 | 0.3 |
| **C–C bonds formed** | | | |
| 0 | 74.5 | 80.4 | 82.1 |
| 1 | 20.5 | 16.8 | 14.5 |
| 2 | 3.7 | 2.1 | 1.9 |
| 3 | 0.7 | 0.5 | 0.8 |
| 4+ | 0.7 | 0.2 | 0.7 |
| **C–X bonds formed** | | | |
| 0 | 59.7 | 39.8 | 46.8 |
| 1 | 30.4 | 51.5 | 45.2 |
| 2 | 8.2 | 7.7 | 6.1 |
| 3 | 1.0 | 0.9 | 1.3 |
| 4+ | 0.6 | 0.2 | 0.5 |

**Table S11**. Summary statistics of the reaction descriptors; *n* is the number of reactions for which the descriptor could be computed.

| Dataset | *n* | Mean | SD | Median | Q1 | Q3 | Min | Max | ≥1 (%) |
|---|---|---|---|---|---|---|---|---|---|
| **Reacting atoms** | | | | | | | | | |
| CbyD | 48,155 | 2.87 | 1.94 | 2 | 2 | 3 | 0 | 52 | 99.14 |
| USPTO-Full | 966,744 | 2.82 | 1.83 | 2 | 2 | 3 | 0 | 34 | 99.98 |
| Pistachio | 4,326,750 | 2.60 | 2.26 | 2 | 2 | 2 | 0 | 63 | 99.92 |
| **Changing rings** | | | | | | | | | |
| CbyD | 48,155 | 0.30 | 0.76 | 0 | 0 | 0 | 0 | 48 | 20.88 |
| USPTO-Full | 966,744 | 0.15 | 0.43 | 0 | 0 | 0 | 0 | 9 | 13.39 |
| Pistachio | 4,326,750 | 0.15 | 0.47 | 0 | 0 | 0 | 0 | 31 | 11.50 |
| **Stereochemical changes (*R/S* + *E/Z*)** | | | | | | | | | |
| CbyD | 48,155 | 0.54 | 1.01 | 0 | 0 | 1 | 0 | 11 | 32.11 |
| USPTO-Full | 966,744 | 0.07 | 0.38 | 0 | 0 | 0 | 0 | 23 | 4.76 |
| Pistachio | 4,326,750 | 0.10 | 0.47 | 0 | 0 | 0 | 0 | 46 | 6.24 |
| **C–C bonds formed** | | | | | | | | | |
| CbyD | 48,155 | 0.34 | 0.81 | 0 | 0 | 1 | 0 | 39 | 25.54 |
| USPTO-Full | 966,744 | 0.23 | 0.53 | 0 | 0 | 0 | 0 | 9 | 19.58 |
| Pistachio | 4,326,750 | 0.24 | 0.64 | 0 | 0 | 0 | 0 | 20 | 17.90 |
| **C–X bonds formed** | | | | | | | | | |
| CbyD | 48,155 | 0.53 | 0.78 | 0 | 0 | 1 | 0 | 15 | 40.26 |
| USPTO-Full | 966,744 | 0.70 | 0.66 | 1 | 0 | 1 | 0 | 8 | 60.22 |
| Pistachio | 4,326,750 | 0.64 | 0.72 | 1 | 0 | 1 | 0 | 14 | 53.20 |

**Table S12**. The ten most frequent transition metals among the reaction agents of each dataset (**Fig. 4C**), with the number of metal-containing agent fragments in parentheses.

| Rank | CbyD | | USPTO-Full | | Pistachio | |
|---|---|---|---|---|---|---|
| | Metal | Count | Metal | Count | Metal | Count |
| 1 | Pd | 3,228 | Pd | 83,115 | Pd | 469,957 |
| 2 | Cu | 1,167 | Cu | 13,310 | Fe | 98,057 |
| 3 | Ru | 786 | Fe | 10,699 | Cu | 63,927 |
| 4 | Zn | 677 | Zn | 7,313 | Zn | 21,379 |
| 5 | Cr | 558 | Ni | 4,255 | Ni | 13,459 |
| 6 | Ti | 512 | Pt | 3,255 | Ti | 9,896 |
| 7 | Ag | 471 | Mn | 2,935 | Pt | 8,841 |
| 8 | Fe | 454 | Ti | 2,447 | Mn | 8,422 |
| 9 | Rh | 430 | Cr | 2,019 | Ag | 6,593 |
| 10 | Ce | 342 | Ag | 1,352 | Cr | 5,974 |

**Table S13**. Route length, longest linear sequence (LLS) and convergence (**Fig. 5**). A route is convergent when its total number of reactions (TNS) exceeds its LLS.

| | **CbyD** | **USPTO-190** | **PaRoutes** |
|---|---|---|---|
| Routes | 3,502 | 190 | 10,000 |
| Convergent routes | 1,351 | 55 | 303 |
| Convergent routes (%) | 38.6 | 28.9 | 3.0 |
| TNS, median | 14 | 6 | 3 |
| TNS, Q1 | 9 | 4 | 2 |
| TNS, Q3 | 21 | 8 | 3 |
| TNS, mean | 16.2 | 6.7 | 2.9 |
| TNS, min | 1 | 2 | 2 |
| TNS, max | 114 | 16 | 10 |
| LLS, median | 13 | 6 | 3 |
| LLS, Q1 | 8 | 4 | 2 |
| LLS, Q3 | 18 | 8 | 3 |
| LLS, mean | 14.1 | 6.0 | 2.9 |
| LLS, min | 1 | 2 | 2 |
| LLS, max | 73 | 14 | 10 |

**Table S14.** Key-step thresholds (**Section S3.4**): a key step is a candidate reaction whose complexity change ΔC is at or above the 90th percentile of all candidates for that score.

| **Score** | **ΔC threshold** | **Key steps** | **Candidates** |
|---|---|---|---|
| Böttcher's score | 34.415 | 3,742 | 38,872 |
| SAScore | 0.658 | 3,888 | 38,872 |
| SPS | 160.000 | 3,961 | 38,872 |
| NPScore | 0.621 | 3,888 | 38,872 |

## S4 INSPECTION OF ROUTES IN OTHER DATASETS

### S4.1 USPTO-190 & PaRoutes

Because USPTO-190 and PaRoutes routes are formed by enumerating single-step reactions, and not from a published experimentally valid synthesis, a sequence can be chemically valid at every step, but invalid as a synthesis route. The route in **Fig. S20A** is one example of this instance. The central intermediate is a secondary alcohol containing an epoxide and a 2,4-difluorophenyl group. That alcohol is acylated as a 3,5-dinitrobenzoate and then hydrolyzed straight back to the alcohol it came from; it is then acylated a second time, as a simple benzoate, and hydrolyzed back again. Each cycle returns the molecule to exactly the compound it started from, so four of the seven steps leave the structure unchanged. Route enumeration can also introduce steps that are chemically reasonable individually but create an unnecessary or misleading sequence in the reconstructed route. One recurring example is a protection step followed immediately by a deprotection step (**Fig. S20B**). In this example, a TBS group is installed on one hydroxyl while the TBS group on a different hydroxyl is removed in the following step.

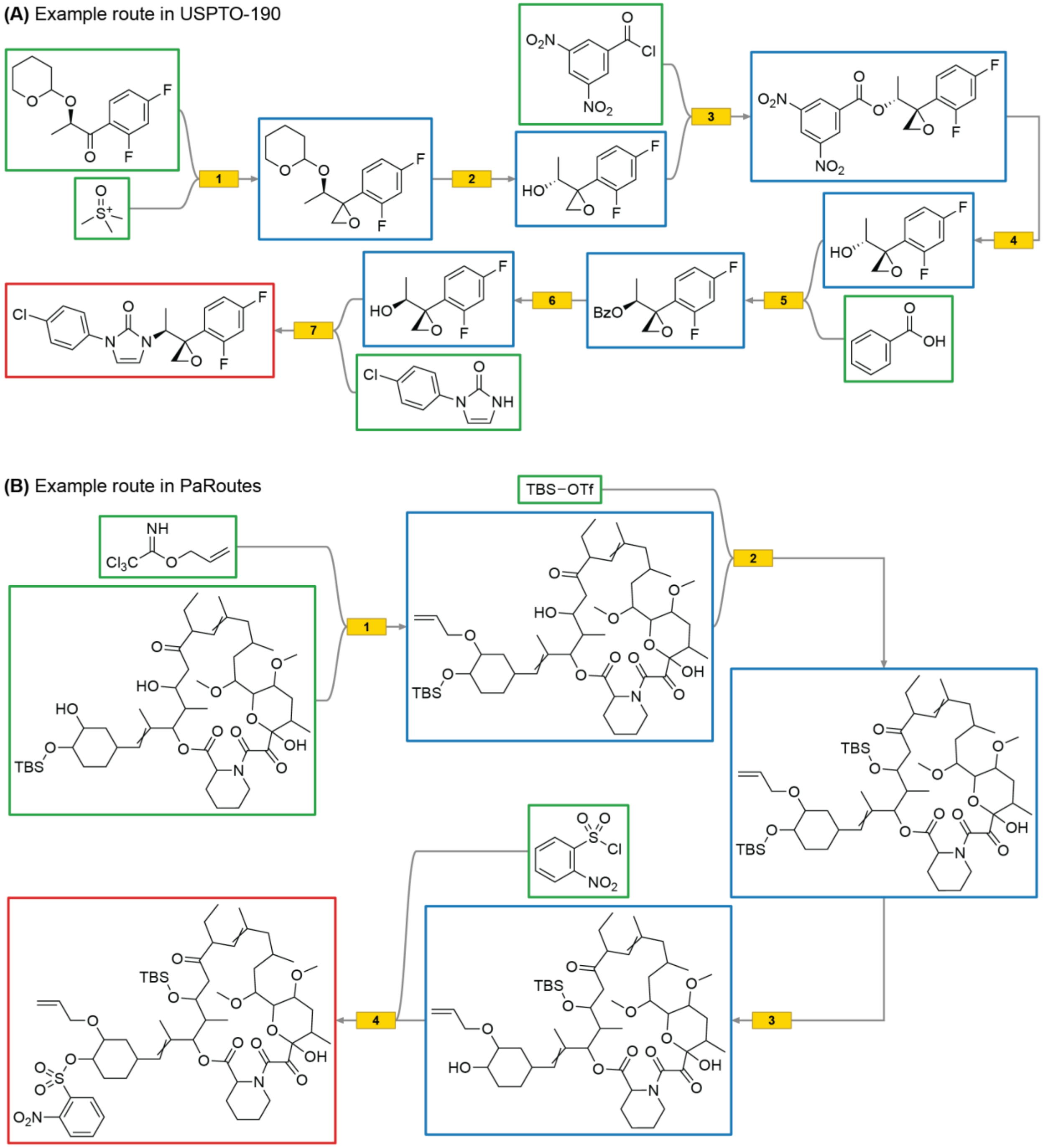

**Figure S20.** Examples of implausible routes in reconstructed route datasets. **(A)** A USPTO-190 route in which a secondary alcohol is esterified and hydrolyzed back twice, returning to the same intermediate before the azole is installed. **(B)** A route in PaRoutes consisting only of protecting-

group manipulation of an advanced intermediate that already contains the complete skeleton of the target.

### S4.2 AbSynth

The two resources, CbyD and AbSynth,[29] share the same goal of making published total syntheses available in machine-readable form. In this section, we compare how a representative route is encoded in each resource and highlight differences in curation and representation.

The most consequential difference lies in the level of reaction annotation. CbyD provides atom-mapped reactions generated using RXNMapper, with reagents corrected and classified, whereas AbSynth records condition information as text. A practical consequence is that reactants contributing atoms to the product are frequently absent from AbSynth's reaction structures and appear only in the condition text (**Fig. S21**). These omitted reactants are often not minor reagents but major structural contributors to the product, as with the amine that forms the benzoxazine ring of hericerin or the tryptamine unit incorporated into eburnamine. Without these reactants, the recorded reactions are unbalanced: atoms in the product have no source, which precludes atom mapping and template extraction. Text-only conditions are also prone to transcription errors, including incorrect stoichiometry, an added reaction temperature, an incorrect reagent formula and a yield misattributed across steps (**Fig. S21A, C, E**).

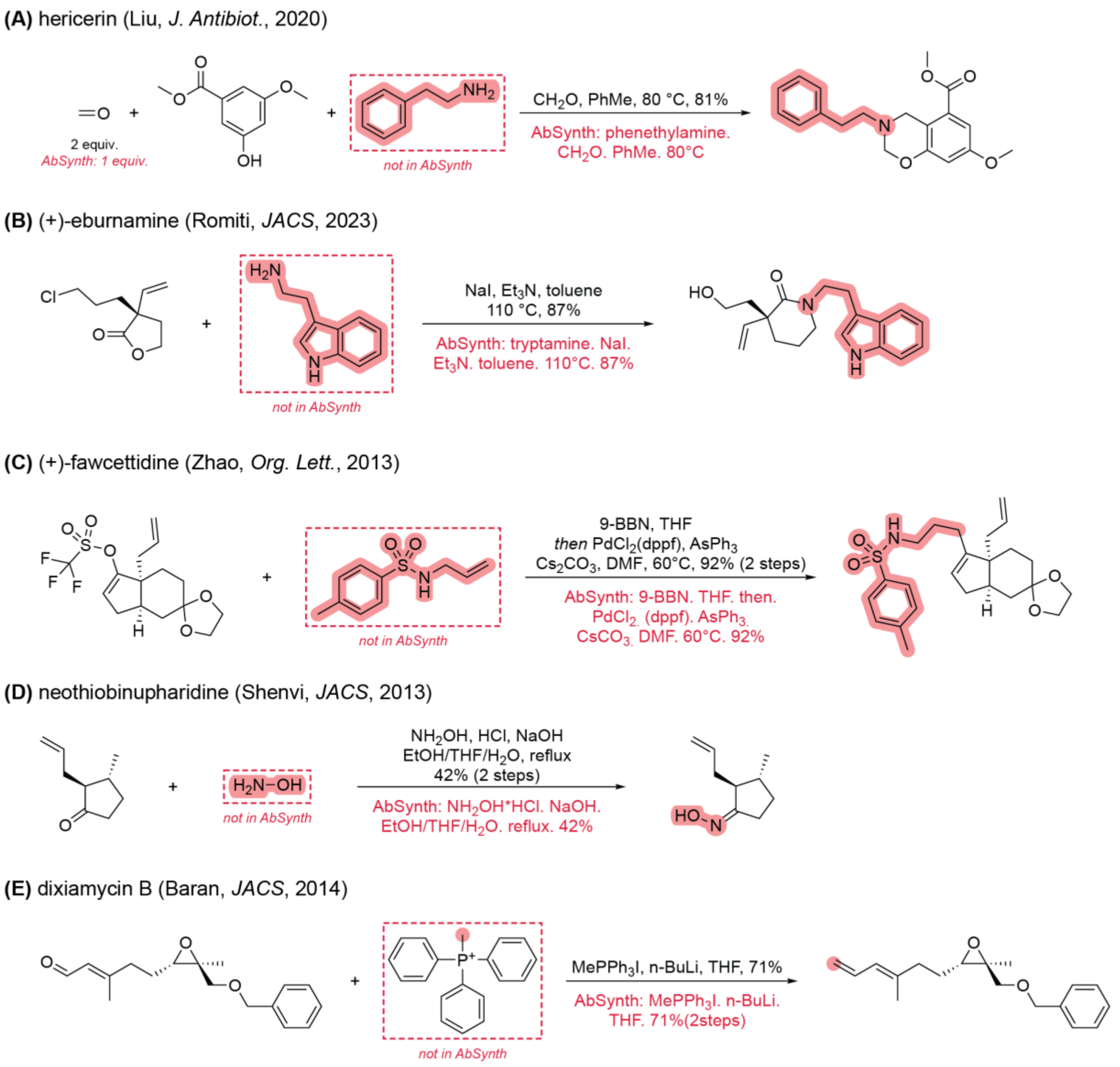

**Figure S21.** Comparison of reaction encoding in CbyD and AbSynth for five representative steps (A–E). Each panel shows the reaction as recorded in CbyD (black), with the corresponding AbSynth condition text in red below the arrow. Reactants in dashed red boxes are missing from AbSynth's reaction structures and appear only as condition text. Red highlighting marks the atoms these reactants contribute to the product, assigned by CbyD atom mapping.

We also compared stereochemical encoding between the two resources. In several reactions, AbSynth records a configuration at a stereocenter that differs from both CbyD and the original publication (**Fig. S22**). Most of these errors fall on spectator stereocenters that should be conserved across the step, such as a cyclopropane center remote from the TBS ether being cleaved in FR-900848 (**Fig. S22A**). In the intramolecular Diels–Alder step toward vinigrol, AbSynth's reactant and product disagree on the configuration of the pivaloyloxy-bearing center, which does not participate in the reaction (**Fig. S22B**). Errors also occur adjacent to the reacting site, as in the misassigned, thio-substituted center of the bicyclomycin diketopiperazine (**Fig. S22C**). Because such errors propagate to downstream intermediates and lie outside the reaction center, they are unlikely to be caught by checks that consider only the transformation itself.

**(A)** FR-900848 (Barrett, *JACS*, 1996)

AbSynth

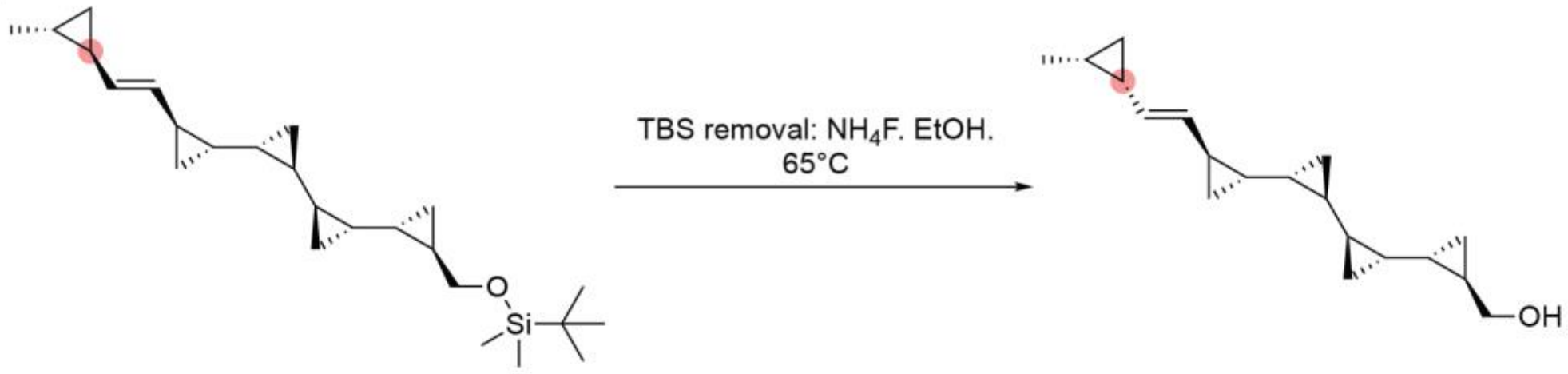

CbyD

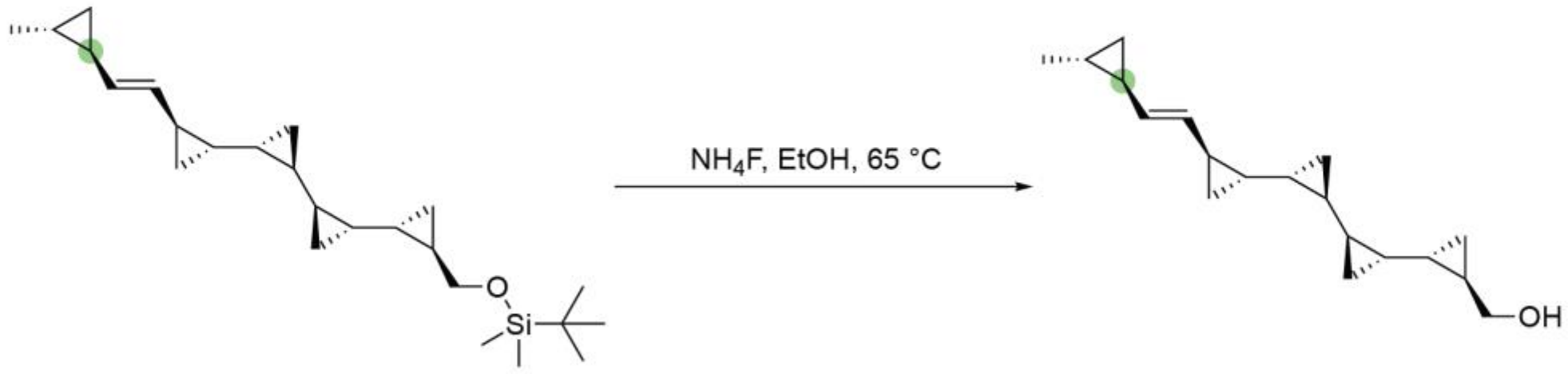

**(B)** vinigrol (Barrett, *Angew. Chem., Int. Ed.*, 2012)

AbSynth

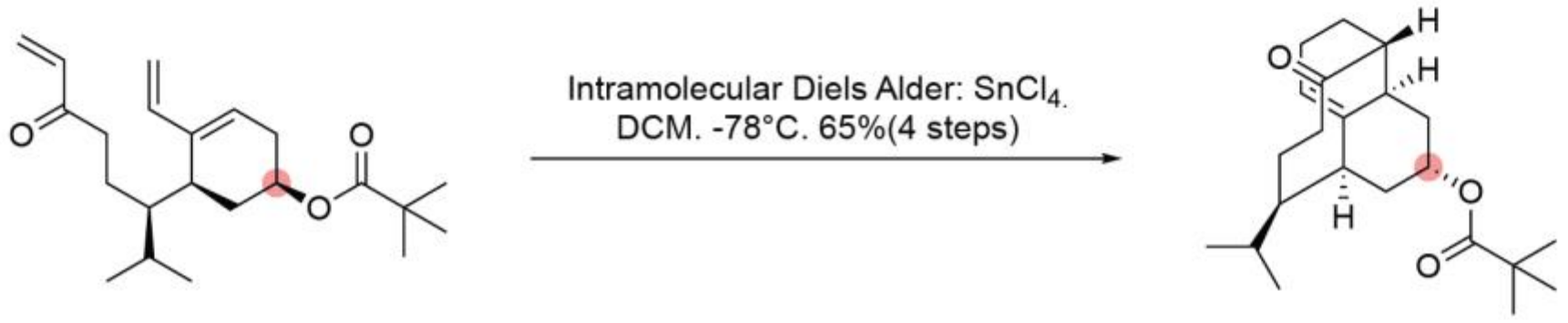

CbyD

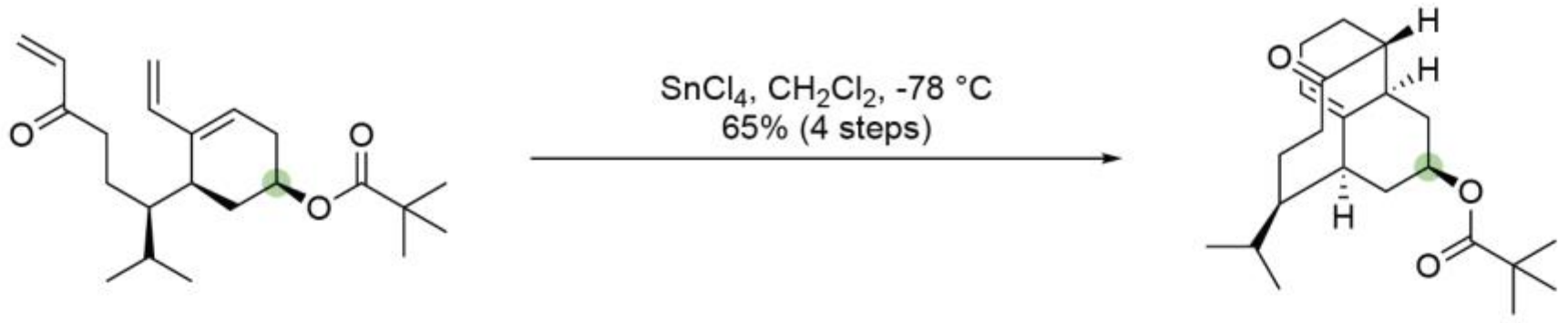

**(C)** bicyclomycin (Williams, *JACS*, 1985)

AbSynth

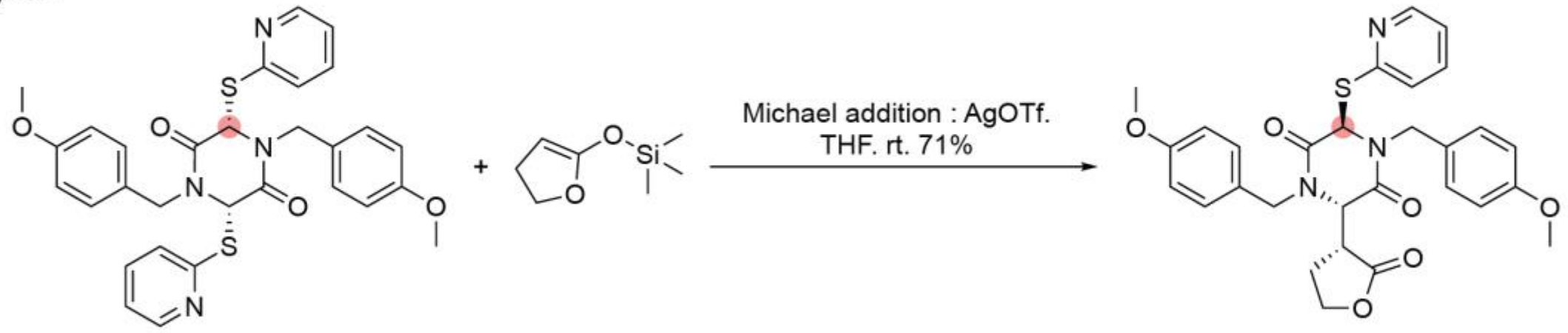

CbyD

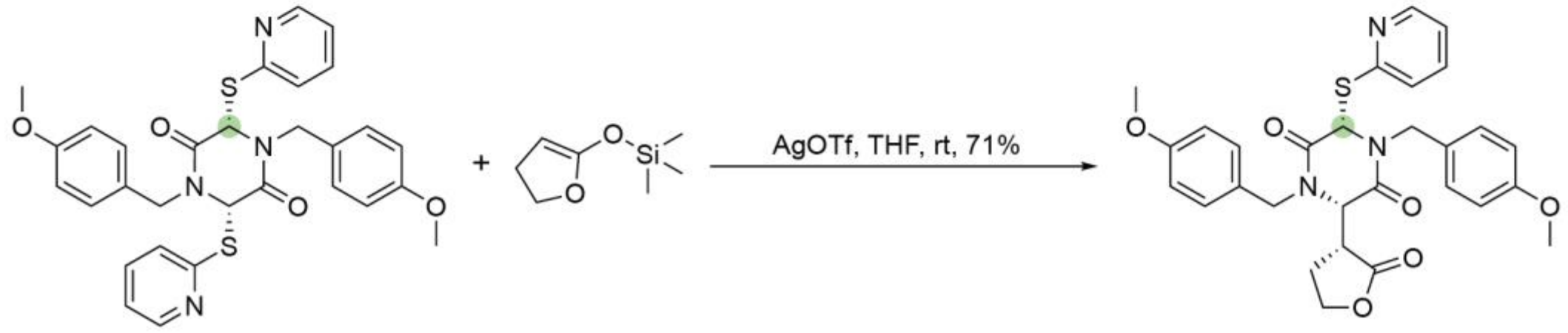

**Figure S22.** Stereochemical discrepancies between AbSynth and CbyD encodings of the same reactions. For each step (A–C), the AbSynth record (top) is shown above the CbyD record (bottom). Highlighted atoms mark stereocenters whose configuration differs between the two resources (red, AbSynth; green, CbyD).

Beyond content, the two resources differ in accessibility. CbyD is openly accessible, versioned with tagged releases, and available for free to anyone on our GitHub repository, whereas AbSynth is accessed through an academic-user registration system and does not currently provide a persistent identifier.